\documentclass[NJP,12pt]{iopart}
\usepackage{iopams}
\usepackage{float}
\usepackage{graphicx}
\usepackage{subfigure}
\usepackage[normalem]{ulem}
\usepackage{cancel}
\usepackage{amssymb}
\usepackage{color}
\usepackage{cite}

\usepackage{bm}
\usepackage{hyperref}
\hypersetup{
colorlinks,
citecolor=blue,
filecolor=blue,
linkcolor=blue,
urlcolor=blue}

\newcommand{\be}{\begin{equation}}
\newcommand{\ee}{\end{equation}} 
\newcommand{\bea}{\begin{eqnarray}}
\newcommand{\eea}{\end{eqnarray}}

\begin{document}

\title[Phenomenology of buoyancy-driven turbulence: recent results]{Phenomenology of buoyancy-driven turbulence: recent results}

\author{Mahendra K. Verma}
\address{Department of Physics, Indian Institute of Technology Kanpur, Kanpur 208016, India}
\ead{mkv@iitk.ac.in}
\author{Abhishek Kumar}
\address{Department of Physics, Indian Institute of Technology Kanpur, Kanpur 208016, India}
\author{Ambrish Pandey}
\address{Department of Physics, Indian Institute of Technology Kanpur, Kanpur 208016, India}
\vspace{10pt}

\begin{abstract}
In this paper, we review the recent developments in the field of buoyancy-driven turbulence.  Scaling and numerical arguments show that the stably-stratified turbulence with moderate stratification has kinetic energy spectrum $E_u(k) \sim k^{-11/5}$ and the kinetic energy flux $\Pi_u(k) \sim k^{-4/5}$, which is called Bolgiano-Obukhov scaling.  The energy flux for the Rayleigh-B\'{e}nard convection (RBC) however is approximately constant in the inertial range that results in  Kolmorogorv's spectrum ($E_u(k) \sim k^{-5/3}$) for the kinetic energy.  The phenomenology of RBC should apply to other flows where the buoyancy feeds the kinetic energy, e.g. bubbly turbulence and fully-developed Rayleigh Taylor instability.  This paper also covers several models that predict the Reynolds and Nusselt numbers of RBC. Recent works show that the viscous dissipation rate of RBC scales as $\sim \mathrm{Ra}^{1.3}$, where $\mathrm{Ra}$ is the Rayleigh number.
\end{abstract}

\tableofcontents

\section{Introduction}
\label{sec:intro}

Gravity pervades the whole universe, and it plays a dominant role in the flow dynamics of the interiors and atmospheres of planets and stars. The gravitational force also affects the engineering flow, e.g., in large turbines.  Therefore,  understanding  the physics of buoyancy-driven turbulence is quite crucial.

Hydrodynamic turbulence is described by Kolmogorov's theory~\cite{Kolmogorov:DANS1941a} according to which the energy spectrum ($E(k)$) in the inertial range is described by
\be
E(k) = K_{Ko} \Pi^{2/3} k^{-2/3},
\ee
where $ K_{Ko} $ is the Kolmogorov's constant, and $\Pi$ is the energy flux or energy cascade rate, which is assumed to be constant in the inertial range.  In Kolmogorov's phenomenology for hydrodynamic turbulence, the flow is forced at large length scales.  However in buoyancy-driven flows, the buoyancy provides forcing at all length scales, hence the kinetic energy flux $\Pi_u$ is expected to be a function of wavenumber $k$. Bolgiano~\cite{Bolgiano:JGR1959} and Obukhov~\cite{Obukhov:DANS1959}   exploited this idea to derive energy spectrum for stably-stratified turbulence; their scaling arguments  yield $\Pi_u(k) \sim k^{-4/5}$, and the kinetic energy spectrum $E_u(k) \sim k^{-11/5}$. Here the kinetic energy is converted to potential energy that leads to decrease of $\Pi(k)$ with $k$.  Procaccia and Zeitak~\cite{Procaccia:PRL1989},  L'vov~\cite{Lvov:PRL1991},  L'vov and Falkovich~\cite{Lvov:PD1992}, and Rubinstein~\cite{Rubinstein:NASA1994} argued that the scaling of Bolgiano~\cite{Bolgiano:JGR1959} and Obukhov~\cite{Obukhov:DANS1959}   would extend to the thermally-driven turbulence as well. Kumar {\em et al.}~\cite{Kumar:PRE2014} however showed that in turbulent convection, the buoyancy feeds the kinetic energy, hence $\Pi_u(k)$ cannot decrease with $k$, and  Bolgiano-Obukhov's arguments are not valid for thermally-driven turbulence.  Using a detailed analysis, Kumar {\em et al.}~\cite{Kumar:PRE2014} showed that turbulent thermal convection shows Kolmogorov's $k^{-5/3}$ energy spectrum.  

Strong gravity makes the flow anisotropic.  Surprisingly the turbulent flow in Rayleigh-B\'{e}nard convection is nearly isotropy~\cite{Nath:arxiv2016}, while the stably-stratified turbulence is close to isotropic when Richardson number is less than unity.  The stably-stratified flows become quasi two-dimensional for larger Richardson numbers.  For Rayleigh-B\'{e}nard convection the large-scale quantities like Reynolds and Nusselt numbers exhibit interesting scaling relations.   

In this short review we describe the recent results of the field. For a more detailed discussion, refer to the review articles~\cite{Ahlers:RMP2009,Bodenschatz:ARFM2000,Lohse:ARFM2010,Siggia:ARFM1994},
We introduce the governing equation and system description in Sec.~\ref{system_description}. We cover recent development on energy spectrum and flux in Sec.~\ref{sec:turbulence}, and scaling of large-scale quantities in Sec.~\ref{sec:large_scales}.   Section~\ref{sec:LSC} contains a brief description of the flow reversal dynamics.  We conclude in Sec.~\ref{sec:summary}.

\section{System description}  
\label{system_description}
In this section we describe the the buoyancy-driven systems  and their associated equations. 

\subsection{Equations under Oberbeck-Boussinesq approximation}
\label{sec:OB}

\begin{figure}[htbp]
\begin{center}
\includegraphics[scale = 0.8]{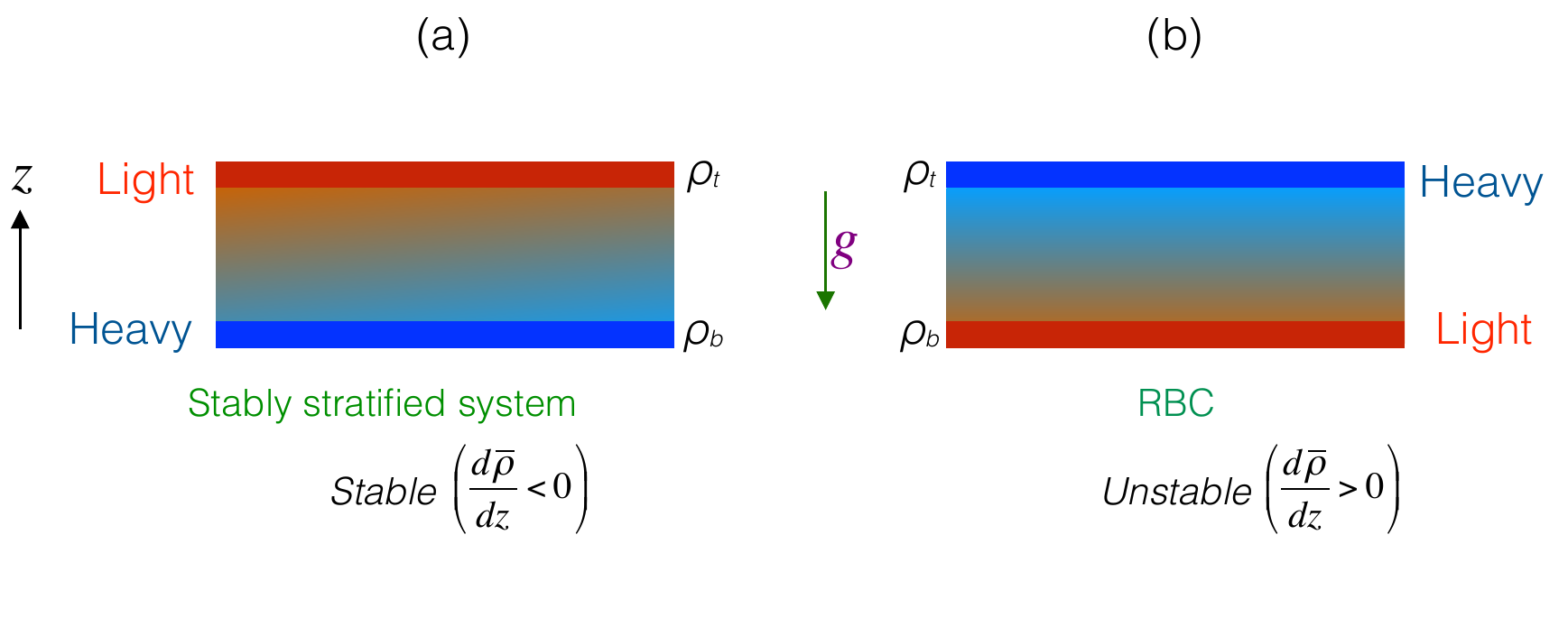}
\end{center}
\setlength{\abovecaptionskip}{0pt}
\caption{Schematic diagrams for the idealized setup of stably stratified  system and Rayleigh-B\'{e}nard convection (RBC): (a) In stably stratified setup, a lighter fluid sits on top of a heavier fluid ($d\bar{\rho}/dz < 0$).  (b) In RBC, heavier (colder) fluid  is on top of lighter (hotter) fluid, thus $d\bar{\rho}/dz > 0$.}
\label{fig:sch_setup}
\end{figure}

Consider fluid between two layers separated by distance $d$ with the bottom density at $\rho_b$ and the top density at $\rho_t$. Clearly the fluid is under the influence of an external density stratification. Under equilibrium condition, the density profile is 
\be
 \bar{\rho}(z) = \rho_b + \frac{d \bar{\rho}}{d z} z  = \rho_b + \frac{\rho_t-\rho_b}{d} z
 \label{eq:rho_bar}
\ee
where $\rho_b, \rho_t$ are the densities at the bottom and top layers respectively (see Fig.~\ref{fig:sch_setup}).   We denote  $ \bar{\rho}(z) $ as the mean density profile.  With fluctuations,  the local density $\rho_l$ (subscript $l$ stands for local) is 
\be
 \rho_l(x,y,z) =   \bar{\rho}+ \rho(x,y,z).
 \label{eq:density}
 \ee
The gravitational force on a unit volume  is $-\rho_l g \hat{z}$, where $-g \hat{z}$ is the acceleration due to gravity.  Hence the gravitational force density on the fluid is
\bea
{\bf F}_g & = &  -g \rho_l \hat{z} = -g (\bar{\rho} + \rho) \hat{z}  = -g \nabla\left( \int^z \bar{\rho}(z') dz'\right) - \rho g \hat{z}.
  \label{eq:Fg}
\eea
The force $\rho g \hat{z}$ occurring due to the change in density from the local value is the {\em buoyancy}.  It is along $-\hat{z}$ for $\rho > 0$ , but along  $\hat{z}$ for $\rho < 0$.

The fluid flow is described by the Navier-Stokes equation
\be
 \rho_l \left[\frac{\partial {\bf u}}{\partial t} + ({\bf u} \cdot \nabla) {\bf u}  \right] =  -\nabla p  + {\bf F}_g+ \mu \nabla^2 \bf u + {\bf f}_u,
 \label{eq:NS0}
 \ee
where ${\bf u},p$ are the velocity and pressure fields respectively,  $\mu$ is the dynamic viscosity of the fluid, and $ {\bf f}_u$ is the external force in addition to the buoyancy.  Substitution of Eq.~(\ref{eq:Fg}) in Eq.~(\ref{eq:NS0}) yields
\be
\rho_l \left[\frac{\partial {\bf u}}{\partial t} + ({\bf u} \cdot \nabla) {\bf u}  \right] =  -\nabla \sigma  -\rho g \hat{z} + \mu \nabla^2 \bf u,
\label{eq:NS1}
\ee
where 
\be
\sigma = p +g   \int^z \bar{\rho}(z') dz' 
\ee
is the modified pressure. 

The continuity equation for the density is
\be
\frac{\partial \rho_l}{\partial t} + \nabla \cdot ( \rho_l {\bf u}) =  \nabla \cdot (\kappa \nabla \rho_l),
\label{eq:continuity0}
\ee
where $\kappa$ is the diffusivity of the density.  We assume that $\kappa$ is constant in space and time.   We can rewrite Eq.~(\ref{eq:continuity0}) as
\be
\nabla \cdot {\bf u} = -\frac{1}{\rho_l} \frac{d \rho_l}{d t}  +\frac{1}{\rho_l}\kappa \nabla^2 \rho_l.
\ee
Now we employ Oberbeck-Boussinesq (OB) approximation according to which  $(d \rho_l/d t)/\rho_l \approx 0$.  Hence the relative magnitude of $\nabla \cdot {\bf u}$ is
\be
\frac{\nabla \cdot {\bf u}}{U/L} \approx \frac{L}{\rho_l U }\kappa \nabla^2 \rho_l \approx \frac{\kappa}{UL} = \frac{1}{\mathrm{Pe}},
\ee
where $L,U$ are the large length and velocity scales respectively, and $\mathrm{Pe}$ is the P\'{e}clet number.  Hence for large $\mathrm{Pe}$, which is often the case for buoyancy-driven flows, we can assume that $\nabla \cdot {\bf u} = 0$.  Therefore, under the Oberbeck-Boussinesq approximation, Eq.~(\ref{eq:continuity0}) gets simplified.  In addition,  we replace $\rho_l $ of Eq.~(\ref{eq:NS1}) with the mean density of the fluid, $\rho_m$. Hence the governing equations for the buoyancy-driven flows are
\bea
 \frac{\partial {\bf u}}{\partial t} + ({\bf u} \cdot \nabla) {\bf u} & =  & -\frac{1}{\rho_m} \nabla \sigma   -\frac{\rho}{\rho_m} g \hat{z} + \nu \nabla^2 \bf u +  {\bf f}_u, \label{eq:NS} \\
\frac{\partial \rho}{\partial t} +  ({\bf u} \cdot \nabla)  \rho & =  &-   \frac{d \bar{\rho}}{d z} u_z +\kappa \nabla^2 \rho,
\label{eq:continuity}
\eea
where $\nu=\mu/\rho_m$ is the kinematic viscosity. The assumption that $\nu,\kappa$ are constants in space and time is also considered to be a part of the OB approximation.  Also note that the buoyancy term, which is a function of variable density, is retained in the Navier-Stokes equation since it is comparable to the other terms of the momentum equation.   In the stably-stratified turbulence, the total energy decays without ${\bf f}_u$, hence, ${\bf f}_u$ is employed to maintain a steady state.

Note that the system is stable when heavy fluid is below the lighter fluid, or $d \bar{\rho}/dz < 0$. Such systems yield wave solution in the linear limit.   On the contrary, when heavy fluid is above the lighter fluid, $d \bar{\rho}/dz > 0$ and the flow becomes unstable and convective. See Schematic diagram of Fig.~\ref{fig:sch_setup} for an illustration.

Temperature field $T$ induces density variation in the following manner:
\be
\rho_l = \rho_b \left[ 1 - \alpha(T-T_b) \right],
\label{eq:rho_T}
\ee
where $\alpha$ is the thermal expansion coefficient, which is assumed to be constant in space and time.  Hence we can rewrite Eqs.~(\ref{eq:NS},\ref{eq:continuity}) in terms of the temperature field.  Let us consider a fluid confined between two thermally-conducting horizontal plates kept at constant temperatures, as shown in Fig.~\ref{fig:sch_setup}(b).  We denote the  temperatures of the bottom and top plates to be $T_b$ and $T_t$ respectively, and $\Delta = T_b - T_t$.

Thermal convection is absent for small $\Delta$.  Under this condition, the temperature profile is linear as
\be
\bar{T}(z) = T_b + \frac{d\bar{T}}{dz} z = T_b - \frac{T_b-T_t}{d} z,
\ee
and the heat is transported by conduction. This configuration has no fluctuation, i.e., ${\bf u} = 0$ and $\rho = 0$.  The flow however becomes unstable and convective when $\Delta$ exceeds a certain critical value.   For such flows it is customary to write the temperature as 
\be
T(x,y,z) = \bar{T}(z) + \theta(x,y,z),
\label{eq:T}
\ee
where $\theta$ is the temperature fluctuation over the background conduction profile $\bar{T}$.  A comparison of Eqs.~(\ref{eq:density},\ref{eq:rho_T},\ref{eq:T}) yields
\be
\rho = -\rho_m \alpha \theta;~~~\frac{d\bar{\rho}}{dz} = -\alpha \frac{d\bar{T}}{dz},
\label{eq:rho_to_theta}
\ee
substitution of which in Eqs.~(\ref{eq:NS},\ref{eq:continuity}) yields the following set of governing equations:
\bea
 \frac{\partial {\bf u}}{\partial t} + ({\bf u} \cdot \nabla) {\bf u} =  -\frac{1}{\rho_m} \nabla \sigma   + \alpha g \theta \hat{z} + \nu \nabla^2 \bf u, \label{eq:NS_RBC}
 \\
\frac{\partial \theta}{\partial t} +  ({\bf u} \cdot \nabla)  \theta = - \frac{d\bar{T}}{dz} u_z +\kappa \nabla^2 \theta,  \label{eq:continuity_RBC} \\
\nabla \cdot {\bf u} =  0,  \label{eq:inc_RBC}
\eea 
The above fluid configuration under OB approximation is called Rayleigh-B\'{e}nard convection (RBC).  For moderate temperature difference, say 30C for water, Onerbeck-Boussinesq approximation is satisfied.  The flow dynamics of RBC is described by Eqs.~(\ref{eq:NS_RBC},\ref{eq:continuity_RBC},\ref{eq:inc_RBC}).

\subsection{Non-Boussinesq flows}
Oberbeck-Boussinesq approximation provides a useful simplification for the analysis of the fluid flow.  Without this approximation, we would need to solve the equations for the velocity, density, and temperature fields.  For an illustration, refer to the set of equations in Sameen {\em et al.}~\cite{Sameen:EPL2009}.  The above description, called {\em non-Boussinsq convection}, is useful in stellar convection where the temperature difference is too large for the OB approximation to  be valid.  This topic, however, is beyond the scope of this paper.

\subsection{Nondimensionalized equations}
Fluid flows are conveniently described by nondimensional equations since they capture relative strengths of various terms of the equations.  Also, they help reduce the number of parameters of the system, which is quite useful for analysis, as well as for the numerical simulations and experiments.  Equations~(\ref{eq:NS},\ref{eq:continuity})  have been nondimensionalized in various ways.  Here, we present two such schemes.   When we use $d$ as the length scale, $\kappa/d$ as the velocity scale, $d^2/\kappa$ as the time scale, and $\Delta \rho = |\rho_b - \rho_t|$ as the density scale,  we obtain the following nondimensional equations:
\bea
\frac{\partial{\mathbf{u}}}{\partial{t}}+ (\mathbf{u}\cdot \nabla)\mathbf{u} & = &  -\nabla\sigma - \mathrm{Ra Pr} \rho \hat{z} +  \mathrm{Pr} \nabla^{2}\mathbf{u},  \label{eq:nondim_NS}\\
\frac{\partial{\rho}}{\partial{t}}+(\mathbf{u}\cdot\nabla)\rho & = & -S  u_{z} + \nabla^{2}\rho, \label{eq:nondim_continuity}
\eea
where $\rho \rightarrow \rho/(\Delta \rho)$, and
\bea
\mathrm{Prandtl \,\, number\,\,}   \mathrm{Pr}  =  \frac{\nu}{\kappa}, \\
\mathrm{Rayleigh \,\, number\,\,}   \mathrm{Ra}  =  \frac{gd^3 \Delta \rho}{\nu \kappa \rho_m},  \\
\mathrm{Normalized \,\, density \,\, gradient\,\,}   S  =  \frac{d}{\Delta \rho} \frac{d \bar{\rho}}{d z}. 
\eea
  For the stably-stratified flows, $S = -1$, but $S=1$ for RBC.   Using Eqs.~(\ref{eq:rho_to_theta}) we can write the above equation in terms of temperature field as follows:
\bea
\frac{\partial{\mathbf{u}}}{\partial{t}}+ (\mathbf{u}\cdot \nabla)\mathbf{u} & = &  -\nabla\sigma + \mathrm{Ra Pr} \theta \hat{z} +  \mathrm{Pr} \nabla^{2}\mathbf{u}  \label{eq:RBC_u_nondim} \\
\frac{\partial{\theta}}{\partial{t}}+(\mathbf{u}\cdot\nabla)\theta  & = & S u_{z} + \nabla^{2}\theta, \label{eq:RBC_theta_nondim}
\eea
for which 
\bea
\mathrm{Ra}  =  \frac{\alpha g \Delta d^3 }{\nu \kappa},
\eea
where $\Delta$ is the temperature difference between the bottom and top plates, as defined earlier. Note however that for large $\mathrm{Ra}$, the aforementioned nondimensional velocity becomes very large ($\sim \sqrt{\mathrm{Ra Pr}}$)~\cite{Grossmann:JFM2000,Verma:PRE2012} that becomes an obstacle for numerical simulations due to very small time-steps.  Hence, in numerical simulations, it is customary to employ $\sqrt{\alpha g \Delta d}$ as the velocity scale, which yields the following set of equations:
\bea
\frac{\partial{\mathbf{u}}}{\partial{t}}+ (\mathbf{u}\cdot \nabla)\mathbf{u} & = &  -\nabla\sigma +  \theta \hat{z} +  \sqrt{\frac{\mathrm{Pr} }{\mathrm{Ra}}} \nabla^{2}\mathbf{u}, \\
\frac{\partial{\theta}}{\partial{t}}+(\mathbf{u}\cdot\nabla)\theta  & = & S u_{z} + \frac{1}{\sqrt{\mathrm{Ra Pr}}} \nabla^{2}\theta.
\eea

For stably-stratified flows, researchers often employ dimensional equations, but with density converted to units of velocity by a transformation~\cite{Lindborg:JFM2008}
\be
b  = \frac{g}{N}\frac{\rho}{\rho_m}
\label{eq:b_def}
 \ee
 where 
 \be
 N = \sqrt{\frac{g}{\rho_m} \left| \frac{d \bar{\rho}}{d z} \right|}
 \ee
is the Brunt-V\"{a}is\"{a}l\"{a} frequency.  In terms of the above variables, the equations become
\bea
\frac{\partial{\mathbf{u}}}{\partial{t}}+ (\mathbf{u}\cdot \nabla)\mathbf{u} & = &  -\nabla\sigma - N b \hat{z} + \nu \nabla^{2}\mathbf{u},  
\label{eq:u_SS} \\
\frac{\partial{b}}{\partial{t}}+(\mathbf{u}\cdot\nabla)b & = & N  u_{z} + \kappa \nabla^{2}b. \label{eq:b_SS}
\eea

The other important nondimensional parameters used for describing the buoyancy-driven flows are
\begin{eqnarray}
\mathrm{Reynolds \,\, number\,\,} \mathrm{Re}  = \frac{u_\mathrm{rms}d}{\nu}, \label{eq:Re} \\
\mathrm{Froude\,\, number\,\, }  \mathrm{Fr}  =  \frac{u_\mathrm{rms}}{d N}, \\\label{eq:Fr}
\mathrm{Richardson\,\, number\,\, }  \mathrm{Ri}  = \frac{1}{\mathrm{Fr}^2}, \label{eq:Ri} 
\end{eqnarray}
where $u_\mathrm{rms}$ is the rms velocity of flow.  Note that the Richardson number is the ratio of the buoyancy and the nonlinearity $(\bf u \cdot \nabla) \bf u$. Another important nondimensional parameter for RBC is the Nusselt number $\mathrm{Nu}$, which is the ratio of the total heat flux (convective plus conductive) and the conductive heat flux, and is computed using the following formula:
\be 
\mathrm{Nu} = \frac{\kappa \Delta/d + \langle u_z \theta \rangle}{\kappa \Delta /d}. \label{eq:Nu_def}
\ee

\subsection{Boundary conditions}
For the velocity field we employ the following set of boundary conditions:
\begin{enumerate}
\item No-slip:  All the components of the velocity field vanish at the walls, i.e., ${\bf u} = 0$.
\item Free-slip: At a wall, the normal component of the velocity field vanishes, i.e., ${\bf u} \cdot \hat{n} = 0$, and the gradient of the parallel components of the velocity vanishes, i.e., $\partial u_\parallel/\partial n = 0$.
\item Periodic: The velocity is periodic, i.e., ${\bf u} ({\bf x} + l L_x\hat{x} + m L_y\hat{y} + n L_z\hat{z}  ) = {\bf u} ({\bf x})$, where $l,m,n$ are integers, and the box is of the size $L_x \times L_y \times L_z$.
\end{enumerate}

For the temperature field, the typical boundary condition used are
\begin{enumerate}
\item Conducting :  Uniform temperature field at the walls, i.e., $\theta = 0$.
\item Insulating: The temperature flux at the wall is zero, i.e., $\partial \theta/\partial n = 0$.
\item Periodic: The temperature fluctuation  is periodic, i.e., $\theta({\bf x} + l L_x\hat{x} + m L_y\hat{y} + n L_z\hat{z}  ) =\theta ({\bf x})$.
\end{enumerate}

\subsection{Exact relations}
 Equations~(\ref{eq:NS},\ref{eq:continuity}) are nonlinear, and hence researchers have not been able to write down general analytic solutions for them.  However, Shraiman and Siggia~\cite{Shraiman:PRA1990} derived the following exact relations for RBC flows:
\begin{eqnarray}
\epsilon_u & = & \frac{\nu^3}{d^4} \mathrm{\frac{(Nu-1)Ra}{Pr^2}}, \label{eq:eps_u}  \\
\epsilon_\theta & = & \kappa \frac{\Delta^2}{d^2} \mathrm{Nu} \label{eq:eps_theta}.
\end{eqnarray}
Also, in  the idealized limit of $\nu=\kappa=0$, using Eqs.~(\ref{eq:u_SS}, \ref{eq:b_SS}), we deduce that the total energy
\begin{eqnarray}
E =  \frac{1}{2} \int \left(u^2 \pm b^2\right) d {\bf r} \label{eq:Energy}
\end{eqnarray}
is conserved for periodic and vanishing boundary conditions.  In the above, the positive sign is for the stably-stratified flow, while the negative sign for the RBC. A  stably-stratified flow is stable, for which the $u^2/2$ and $b^2/2$ terms are the the kinetic and potential energies of the system, analogous to a harmonic oscillator.   In RBC, the conserved quantity is also written as $\int [ u^2 - \alpha g \theta^2/(d\bar{T}/dz) ] /2 d {\bf r}$, where $\theta^2/2$ is called {\em entropy}. Note that $\theta^2/2$  is not the thermodynamic entropy that quantifies the degree of disorder at the microscopic scales.

It is  convenient to describe behaviour of turbulence flows in spectral or Fourier space since it captures the scale-by-scale energy and interactions quite well.  In the next subsection, we describe the definitions used for such descriptions.

\subsection{Equations in Fourier space}

We rewrite Eqs.~(\ref{eq:NS_RBC})-(\ref{eq:inc_RBC}) in the Fourier space as
\begin{eqnarray}
\left(\frac{d}{dt} + \nu k^2 \right) {u}_i ({\bf k}, t ) &  = &  - i k_i \frac{{\sigma} ({\bf k},t)}{\rho_m} -i k_j \sum_{{\bf k=p+q}}{u}_j ({\bf q}, t ) {u}_i ({\bf p}, t )  \\ \nonumber  
& & + \alpha g {\theta}({\bf k}, t) \hat{z} + \nu k^2 u_i({\bf k},t), \label{eq:NSk}\\
\left(\frac{d}{dt} +\kappa k^2 \right){\theta} ({\bf k}, t ) &  = & -\frac{d\bar{T}}{dz} \hat{u}_z({\bf k},t) -i k_j \sum_{{\bf k=p+q}}{u}_j ({\bf q}, t ) {\theta} ({\bf p}, t ),\\ 
 k_i {u}_i ({\bf k},t)  & = & 0. \label{eq:continuity2}
\end{eqnarray}
In the above equations, $i$ represents two things: $\sqrt{-1}$ in front of the $k_i {{p} ({\bf k},t)}/{\rho_m}$ term, and $i=x,y,z$ in ${u}_i$. Note that ${\bf {u}(k)}$, ${p} ({\bf k})$, and ${\theta} ({\bf k})$ are the Fourier transforms of ${\bf u}$, $p$, and $\theta$ respectively.  The above equations are in terms of $\theta$, but we can easily convert them as a function of $\rho$.

In the Fourier space, $E_u(k)$  denotes the kinetic energy spectrum, which is the sum of the kinetic energy of all the modes in a given shell $(k-1, k]$.  Similarly we define the spectra for the entropy and potential energy, which are denoted by $E_\theta$ and  $E_b$ respectively.   They are computed using the following formulas:
\begin{eqnarray}
E_u(k) &=& \sum_{k -1 < k^{\prime} \leq k} \frac{1}{2} |{\bf {u}}({\bf k^\prime})|^2, \label{eq:KE_spectrum} \\
E_{\theta}(k) &=& \sum_{k -1 < k^{\prime} \leq k} \frac{1}{2} |{\theta}({\bf k^\prime})|^2. \label{eq:entropy_spectrum}  \\
E_b(k) &=& \sum_{k -1 < k^{\prime} \leq k} \frac{1}{2} |b({\bf k^\prime})|^2. \label{eq:PE_spectrum} 
\end{eqnarray}

\subsection{Linear and nonlinear regimes}
The  behavior of buoyancy driven flows depends on the  parameters and dimensionality.  Here we present a bird's-eye view of the observed states of RBC and stably-stratified flows.

\subsubsection{RBC}
It can be easily shown that Eqs.~(\ref{eq:RBC_u_nondim},\ref{eq:RBC_theta_nondim}) yield a unstable solution at $ \mathrm{Ra} = \mathrm{Ra}_c$, with $\mathrm{Ra}_c = 27 \pi^4/4$ for the free-slip boundary conditions, and $\mathrm{Ra}_c \approx 1708$ for the no-slip boundary condition~\cite{Chandrasekhar:book:Instability}.  The unstable solutions are the convective rolls.   For $\mathrm{Ra}$ just above the onset, the instability saturates due to nonlinearity leading to  the ``roll" solutions.  At larger $\mathrm{Ra}$, the nonlinearity yields patterns and chaos~\cite{Chandrasekhar:book:Instability,Manneville:book:Instabilities,Busse:PRL1978,Pal:EPL2009,Mishra:EPL2010,Bhattacharjee:Book}.  For even larger nonlinearity, spatio-temporal chaos, weak turbulence, and strong turbulence emerge~\cite{Manneville:book:Instabilities}.  In this paper we will focus on only the strong turbulence regime.

\subsubsection{Stably-stratified flow}  For $S=-1$, the linearised version of Eqs.~(\ref{eq:nondim_NS}, \ref{eq:nondim_continuity}) yields  internal gravity waves whose dispersion relation is
\be
\omega = \frac{k_\perp}{k} N, \label{eq:N_freq}
\ee
where $k_\perp = \sqrt{k_x^2 + k_y^2}$ is the wavenumber component perpendicular to the buoyancy direction.   Clearly $\omega = N$ for $k_\parallel = 0$.  These internal gravity waves persist for weak nonlinearity and inviscid case ($\nu=\kappa=0$).  Strong nonlinearity has two kinds of generic behaviour: Strong stratification ($\mathrm{Fr} \ll 1$) suppresses the flow along the buoyancy direction and yields a quasi two-dimensional (2D) stratified flow; on the other hand, moderate and weak stratification ($\mathrm{Fr} \gtrapprox 1$) yields near isotropic turbulent flows.  For $\mathrm{Fr} \approx 1$, Kumar {\em et al.}~\cite{Kumar:PRE2014} obtained Bolgiano-Obukhov~\cite{Bolgiano:JGR1959,Obukhov:DANS1959} scaling as predicted (to be described in Sec.~\ref{sec:SSF}).   In this paper we focus on the $\mathrm{Fr} \gtrapprox 1$ regime.

\subsection{Temperature profile and related equations}
\label{subsec:T(z)}

In this subsection we derive the properties of temperature fluctuations.  For convenience we work with nondimensional variables.

\begin{figure}
\begin{center}
\includegraphics[scale=1]{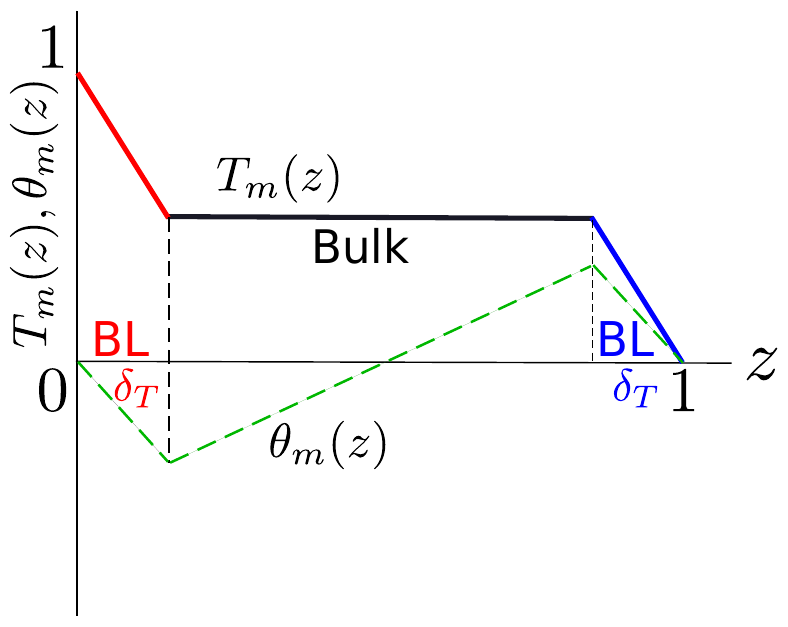}
\caption{A schematic diagram of the planar-averaged temperature  as a function of the vertical coordinate. The temperature drops sharply to $1/2$ in the boundary layers. Reprinted with permission from Pandey and Verma~\cite{Pandey:PF2016}.}
\label{fig:Tprofile}
\end{center}
\end{figure}

Experiments and numerical simulations of RBC reveal that the horizontally averaged temperature $T_m(z)$ remains approximately a constant ($\approx 1/2$) in the bulk, and its value drops sharply in the thermal boundary layers~\cite{Emran:JFM2008, Shishkina:JFM2009}, as shown in Fig.~\ref{fig:Tprofile}.  The quantitative expression for $T_m(z) = \langle T \rangle_{xy}$ can be approximated as
\begin{equation}
T_m(z)= 
\cases{
    1 - \frac{z}{2\delta_T} & {for} $0 < z < \delta_T$ \\
	1/2  & {for} $\delta_T < z < 1 - \delta_T$ \\
	\frac{1-z}{2\delta_T} &{for} $1 - \delta_T < z < 1$ \\}
\end{equation}
where $\delta_T$ is the thickness of the thermal boundary layer, and $\langle \rangle_{xy}$ represents averaging over the $xy$ planes.  A horizontal averaging of Eq.~(\ref{eq:T}) yields $\theta_m(z) = T_m(z) + z-1$, and hence $\theta_m(z)$ is
\begin{equation}
    \theta_m(z) = 
\cases{
    z \left( 1 - \frac{1}{2\delta_T} \right ) & {for} $0 < z < \delta_T$ \\
	z - 1/2 & {for} $\delta_T < z < 1 - \delta_T$ \\
    (z-1)\left( 1 - \frac{1}{2\delta_T} \right ) & {for} $1 - \delta_T < z < 1$ 
}
\end{equation}
as exhibited in Fig.~\ref{fig:Tprofile}.  For thin thermal boundary layers, ${\theta}_m(0,0,k_z)$, which is the Fourier transform of $\theta_m(z)$, is dominated by the contributions from the bulk.  Hence

\begin{eqnarray}
{\theta}_m(0,0,k_z) & = & \int_0^1 \theta_m(z) \sin(k_z \pi z) dz \nonumber \\
 & \approx & \int_0^1 (z-1/2) \sin(k_z \pi z) dz  \nonumber\\
 & \approx &\cases{
       -\frac{1}{\pi k_z} &$\mathrm{for \, even} \, k_z$ \\
        0 \, &$\mathrm{otherwise}$
}
\label{eq:theta_hat_k}
\end{eqnarray}

The corresponding velocity mode,  ${u}_z(0, 0, k_z) = 0$ because of the incompressibility condition ${\bf k} \cdot {\bf u}(0, 0, k_z) = k_z {u}_z(0,0,k_z) = 0$.  Also, ${u}_x(0, 0, k_z) = {u}_y(0, 0, k_z) =  0$ in the absence of a  mean flow along the horizontal direction.  Hence for the ${\bf k} = (0,0,k_z)$ modes, the momentum equation yields
\begin{equation}
0  = -\frac{ i {\bf k} {\sigma}({\bf k})}{\rho_0} + \alpha g {\theta}({\bf k}) \hat{\bf z} 
\end{equation}
or $d \sigma_m(z) /dz = \rho_0 \alpha g \theta_m$, and the dynamics of the remaining set of Fourier modes is governed by the momentum equation as
\begin{equation}
\frac{\partial {{\bf u}}({\bf k})}{\partial t} + i \sum_{\bf{p+q=k}} {[\bf k \cdot {u}}({\bf q})] {\bf {u}}({\bf p})  = -\frac{  i {\bf k} {\sigma}_{\mathrm{res}}({\bf k})}{\rho_0} + \alpha g {\theta}_\mathrm{res}({\bf k}) {\bf z} - \nu k^2 \hat{{\bf u}}({\bf k}), \label{eq:u_four}
\end{equation}
where
\begin{eqnarray}
\theta = \theta_\mathrm{res} + \theta_m;~~~  \sigma & = & \sigma_\mathrm{res} + \sigma_m. \label{eq:theta_res} 
\end{eqnarray}
Hence,  the modes ${\theta}_m(0,0, k_z)$ and ${\sigma}_m(0,0, k_z)$ do not couple with the velocity modes in the momentum equation, but $\theta_\mathrm{res}$ and $\sigma_\mathrm{res}$ do.  

Equation~(\ref{eq:u_four})§ has strong implications on the scaling of the Reynolds and Nusselt numbers, which will be discussed in Sec.~\ref{sec:large_scales}.  In addition, the set of Fourier modes $\theta(0,0,k_z)$ of Eq.~(\ref{eq:theta_hat_k}) yields $E_\theta(k) \sim k^{-2}$.  This issue will be discussed in Sec.~\ref{sec:turbulence}.

\subsection{Other related systems}
Several buoyancy-driven systems can be related to RBC.  Here we list  some of these systems.

\subsubsection{Rayleigh-Taylor instability (RTI)}  A fluid configuration with a denser fluid above a lighter fluid is unstable.  The heavier fluid falls and the lighter fluid rises.  After an initial stage of RTI, the flow develops significant nonlinearity and becomes turbulent~\cite{Chertkov:PRL2003}.   We will discuss later that the turbulence phenomenology of RTI is similar to that of RBC.

\subsubsection{Taylor-Couette flow} Two coaxial rotating cylinders create random flow at large Taylor number.  This flow has been related to RBC with significant similarities in their phenomenology.  See Grossmann {\em et al.}~\cite{Grossmann:ARFM2016} for a review of such flows.

\subsubsection{Turbulent exchange flow in a vertical pipe}~\label{subsec:Arakeri} Arakeri and coworkers~\cite{Arakeri:CurrSci2000}  performed experiments in which a flow  in a vertical tube is driven by an unstable density difference across the tube.  They placed a brine solution at the top and distilled water at the bottom. This system has significant similarities with RBC~\cite{Arakeri:CurrSci2000}. Note however that the above system does not have walls or boundary layers at the top and bottom; this feature helps us study the ultimate regime quite conveniently.  Exchanging the top and bottom containers will lead to  behaviour similar to stably-stratified flows.

\subsubsection{Bubbly flow} \label{subsec:bubbly_flow} Bubbles are introduced in a tank in which turbulence is generated by an active grid~\cite{Prakash:JFM2016}.  Naturally this system has certain similarities with RBC. 

In the next section, we will relate the turbulence behaviour of the above systems.

\section{Spectra and fluxes of buoyancy-driven turbulence}
\label{sec:turbulence}

It is  convenient to describe behaviour of turbulence flows in spectral or Fourier space since it captures the scale-by-scale energy and interactions very well.  In the next subsection, we describe the definitions used for such description.

\subsection{Definitions}

We can derive the time-evolution equation for $E_u(k)$ using Eq.~(\ref{eq:NS}) as~\cite{Lesieur:book,Verma:EPL2012}
\begin{equation}
\frac{\partial E_u(k)}{\partial t} = T_u(k) + F_B(k) + F_\mathrm{ext}(k) - D(k),
\label{eq:dEk_dt}
\end{equation}
where $T_u(k)$ is the energy transfer rate to the shell $k$ due to nonlinear interaction, and $F_B(k)$ and $F_\mathrm{ext}(k)$ are  the energy supply rates to the shell from the  buoyancy  and external forcing ${\bf f}_u$ respectively, i.e.,
 \begin{eqnarray}
F_B(k) & = & - \sum_{|{\mathbf k}| = k}  g \Re\langle u_z({\mathbf k}) \rho^*({\mathbf k}) \rangle, \\ \label{eq:FB} 
F_\mathrm{ext}(k) & = & \sum_{|{\mathbf k}| = k}  \Re\langle {\mathbf u}({\mathbf k}) \cdot {\mathbf f_u}^*({\mathbf k}) \rangle. \label{eq:Fext}
 \end{eqnarray}
For brevity we set $\rho_m =1 $.   In Eq.~(\ref{eq:dEk_dt}), $D(k)$ is the viscous dissipation rate  defined by
 \begin{equation}
D(k) = \sum_{|{\mathbf k}| = k} 2 \nu k^2 E_u(k).
\label{eq:D_k}
\end{equation}

The kinetic energy (KE) flux $\Pi_u(k_0)$,  which is defined as the kinetic energy leaving a wavenumber sphere of radius $k_0$ due to nonlinear interactions,  is related to the nonlinear interaction term $T_u(k)$ as 
\begin{equation}
\Pi_u(k) = - \int_0^k T_u(k)dk.
\label{eq:Pi_def}
\end{equation}
Under a steady state ($\partial E_u(k)/\partial t = 0$), using Eqs.~(\ref{eq:dEk_dt}) and (\ref{eq:Pi_def}), we deduce that
\begin{equation}
\frac{d}{dk} \Pi_u(k) =  F_B(k) + F_{ext}(k) - D(k)
\label{eq:dPik_dk}
\end{equation}
or
\begin{equation}
\Pi_u(k+\Delta k) =  \Pi_u(k) + [ F_B(k) + F_{ext}(k) - D(k)] \Delta k.
\label{eq:Pi_k_Fk_Dk}
\end{equation}

In computer simulations, the KE flux, $\Pi_u(k_0)$,  is computed using the following formula~\cite{Dar:PD2001,Verma:PR2004},
\begin{equation}
\Pi_u(k_0)  =  \sum_{k > k_0} \sum_{p\leq k_0} \delta_{\bf k,\bf p+ \bf q} \Im([{\bf k \cdot u(q)}]  [{\bf u^*(k) \cdot u(p)}]). \label{eq:ke_flux}
\end{equation}
Similarly, the potential energy (PE) flux $\Pi_{\rho}(k_0)$ is the potential energy leaving a wavenumber sphere of radius $k_0$, which is computed using 
\begin{eqnarray}
\Pi_{\rho}(k_0)  & = &  \sum_{k > k_0} \sum_{p \leq k_0} \delta_{\bf k,\bf p+ \bf q} \Im([{\bf k \cdot u(q)}]  [{ b^*({\bf k}) b({\bf p})}]), \label{eq:pe_flux} 
\end{eqnarray}
where $b$ is defined in Eq.~(\ref{eq:b_def}).  For RBC, we replace ${\bf u}$ and $b$ by nondimensional ${\bf u}$ and $\theta$ respectively. 

For a more detailed description of the energy transfers, we divide the wavenumber space into a set of wavenumber shells.  The energy contents of a wavenumber shell of radius $k$ and of unit width is denoted by $E(k)$.  The shell-to-shell energy transfer rate from the velocity field of the $m$th shell to the velocity field of the $n$th shell is defined as
\begin{equation}
T_{n}^{m}=\sum\limits _{\mathbf{k}\in n}\sum\limits _{\mathbf{p}\in m} \delta_{\bf k,\bf p+ \bf q} \Im([{\bf k \cdot u(q)}]  [{\bf u^*(k) \cdot u(p)}]).
\label{eq:S2S}
\end{equation}

 One of the most interesting problems in the field of buoyancy driven turbulence is the scaling of energy spectra and fluxes~\cite{Lohse:ARFM2010,Riley:bookTen}. In the next section, we will review some of the  theoretical results obtained for the aforementioned topic.

\subsection{Turbulence phenomenology}

\subsubsection{Classical Bolgiano-Obukhov scaling for stably-stratified turbulence (SST):}
\label{subsec:BO_intro}
For the inertial range of isotropic hydrodynamic turbulence, Kolmogorov~\cite{Kolmogorov:DANS1941a} first proposed a phenomenology according to which the energy spectrum is independent of the fluid properties and nature of large-scale forcing. He showed that the one-dimensional energy spectrum $E(k)=K_{Ko}\Pi_u^{2/3}k^{-5/3}$ in the inertial range of wavenumbers, where $\Pi_u(k)$ is the constant energy flux, and $K_{Ko}$ is the Kolmogorov's constant.

Buoyancy (forcing) act at all scales, hence Kolmogorov's theory may not work for the buoyancy-driven turbulence.  In this section we will describe how the buoyancy affects the energy spectra and fluxes of the buoyancy-driven flows.  For stable stratification, Bolgiano~\cite{Bolgiano:JGR1959} and Obukhov~\cite{Obukhov:DANS1959} argued that the KE flux $\Pi_u(k)$ is depleted at different length scales due to the conversion of  KE to  PE via buoyancy ($u_z \rho g$). Subsequently, $\Pi_u(k)$ decreases with $k$, and $E_u(k)$ is steeper than that predicted by Kolmogorov's theory;  we refer to the above as BO phenomenology or scaling. According to this phenomenology, for $L_B \ll l \ll L$,  buoyancy is important and the buoyancy term is balanced by the nonlinear term [$\rho g  \approx (\bf u \cdot \nabla) \bf u$].  Here $L_B$ is the Bolgiano scale~\cite{Bolgiano:JGR1959} and $L$ is the large length scale or  the box size.  The force balance  at wavenumber  $k = 1/l$ yields
\begin{equation}
\rho_k g \approx k u_k^2  . \label{eq:u_l}
\end{equation}
  According to BO phenomenology,  PE has a constant flux, i.e., $\Pi_{\rho} \approx k u_k \rho_k^2 \approx \epsilon_{\rho} $.  Hence,  \begin{eqnarray}
u_k & \approx &   \epsilon_{\rho}^{1/5} g^{2/5} k^{-3/5}, \label{eq:u_k} \\
\rho_k & \approx &   \epsilon_{\rho}^{2/5} g^{-1/5} k^{-1/5}. \label{eq:theta_k} 
\end{eqnarray}
Therefore, the KE spectrum $E_u(k) \approx u_k^2/k$, PE spectrum $E_{\rho}(k) \approx \rho_k^2/k$, and $\Pi_u(k) \approx u_k^3 k$  are
\begin{eqnarray}
E_u(k) & =  & c_1 \epsilon_\rho^{2/5}  g^{4/5} k^{-11/5}, \label{eq:Eu} \\
E_\rho(k) & =  & c_2 \epsilon_\rho^{4/5}  g^{-2/5} k^{-7/5}, \label{eq:Etheta} \\
\Pi_u(k) & = & c_3  \epsilon_\rho^{3/5} g^{6/5} k^{-4/5},  \label{eq:pi} \\
\Pi_\rho(k) & = &  \epsilon_\rho, \label{eq:pi_theta} 
\end{eqnarray}
where $c_i$'s are constants.  At smaller length scales ($k>k_B$), where $k_B = 1/L_B$ is the Bolgiano wavenumber,  Bolgiano~\cite{Bolgiano:JGR1959} and Obukhov~\cite{Obukhov:DANS1959} argued that the buoyancy is relatively weak, hence Kolmogorov-Obukhov (KO) scaling is valid in this regime, i.e.,
\begin{eqnarray}
E_u(k) & =  & K_{Ko}  \epsilon_u^{2/3}k^{-5/3}, \label{eq:Eu_KO} \\
E_\rho(k) & =  & K_{Ba} \epsilon_u^{-1/3}\epsilon_\rho k^{-5/3}, \label{eq:Etheta_KO} \\
\Pi_u(k) & = &  \epsilon_u,  \label{eq:pi_KO} \\
\Pi_\rho(k) & = &  \epsilon_\rho, \label{eq:pi_theta_KO}
\end{eqnarray}
where $K_{Ba}$ is the Batchelor's constant. A comparison of $\Pi_u(k)$ of Eq.~(\ref{eq:pi}) with that of Eq.~(\ref{eq:pi_KO}) yields the crossover wavenumber $k_B$ as
\begin{equation}
k_B \approx  g^{3/2} \epsilon_u^{-5/4} \epsilon_{\rho}^{3/4}. \label{eq:kb}
\end{equation}

The BO phenomenology implicitly assumes isotropy in Fourier space, which is a tricky assumption.  For BO scaling, the gravity must be strong enough to compete with the nonlinear term ${\bf u}\cdot \nabla {\bf u}$, but not too strong to make the flow quasi two-dimensional (quasi-2D). This corresponds to $\mathrm{Fr} \approx 1$ regime.  A large number of earlier explorations in SST have been for $\mathrm{Fr} \ll 1$ regime, see for example, Lindborg~\cite{Lindborg:JFM2006}, Brethouwer~\cite{Brethouwer:JFM2007}, and Bartello and Tobias~\cite{Bartello:JFM2013}. SST can be broadly classified in three regimes.  Note that  nonlinearity is strong  ($\mathrm{Re} \gg 1$) for turbulent flows.
\begin{enumerate}
\item Weak gravity ($\mathrm{Ri} \ll 1$):  Strong nonlinearity yields behaviour similar to hydrodynamic turbulence ($E_u(k) \sim k^{-5/3}$).

\item Moderate gravity ($\mathrm{Ri} \approx 1$): Comparable strengths of gravity and nonlinearity yields nearly isotropic turbulence with BO scaling, as described earlier.

\item Strong gravity ($\mathrm{Ri} \gg 1$): Strong gravity makes the flow quasi-2D.  Hence the behaviour has similarities with 2D hydrodynamic turbulence (e.g., inverse cascade of energy).  Refer to Lindborg~\cite{Lindborg:JFM2006}, Brethouwer~\cite{Brethouwer:JFM2007}, and Bartello and Tobias~\cite{Bartello:JFM2013} for further details.
\end{enumerate}

\subsubsection{Generalization of Bolgiano-Obukhov scaling to RBC:}
\label{sub:BO_GEN}

Using mean field theory approximation, Procaccia and Zeitak~\cite{Procaccia:PRL1989} argued that the Bolgiano-Obukhov  scaling  is applicable  to  convective  turbulence. Later,  L'vov~\cite{Lvov:PRL1991} assumed that in convective turbulence, the kinetic energy is converted to the potential energy and therefore, favored  BO scaling.   L'vov and Falkovich~\cite{Lvov:PD1992} employed a differential model for energy and entropy fluxes in $k$-space and found that the BO scaling is valid for convective turbulence.  Rubinstein~\cite{Rubinstein:NASA1994} employed  renormalization group analysis to RBC and observed that the renormalized viscosity $\nu(k) \sim k^{-8/5}$, $E_u(k) \sim k^{-11/5}$, and $E_{\rho}(k) \sim k^{-7/5}$.  Based on these observations Rubinstein claimed  BO scaling for RBC.

The aforementioned theories had profound influence in the field, and a large number of analytical, experimental, and numerical works attempted to verify these ideas.  Recently Kumar {\em et al.}~\cite{Kumar:PRE2014} showed that the BO scaling does not describe RBC turbulence since the energy supply by buoyancy in RBC is very different from that in stably stratified flow.  We will provide these arguments below.

\subsubsection{A phenomenological argument based on kinetic energy flux:}
\label{sec:pheno}

Kumar {\em et al.}~\cite{Kumar:PRE2014} and Verma {\em et al.}~\cite{Verma:IUTAM2015,Verma:PhysFocus2015} presented a phenomenological argument based on the KE flux to derive a spectral theory of buoyancy-driven turbulence.  Equation~(\ref{eq:Pi_k_Fk_Dk}) provides   important clues on the energy spectrum and flux of the buoyancy-driven flows.  Here we list three possibilities for the inertial range ($k_f < k < k_d$), where $k_f$ is the forcing wavenumber, and $k_d$ is the dissipation wavenumber:
\begin{enumerate}
\item  For the inertial range of hydrodynamic turbulence, $F_B(k)=0$  and $D(k) \rightarrow 0$, therefore \, $\Pi_u(k+\Delta k) \approx  \Pi_u(k)$ and $E_u(k) \sim k^{-5/3}$, which is a prediction of the Kolmogorov's theory~\cite{Kolmogorov:DANS1941a}. Note that $F_{ext}(k)=0$ in the inertial range.
\label{enu_1}

\begin{figure}[htbp]
\begin{center}
\includegraphics[scale = 1.1]{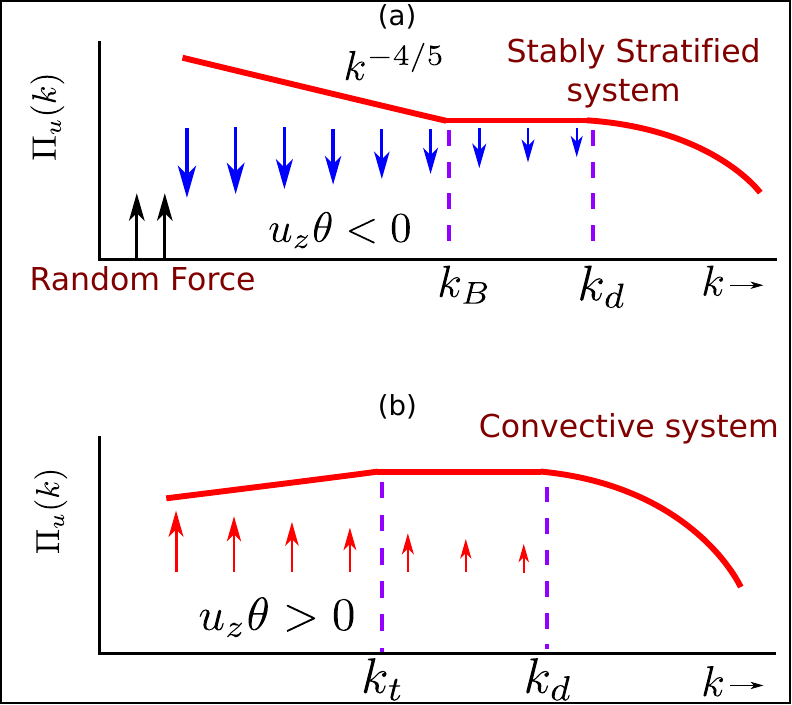}
\end{center}
\setlength{\abovecaptionskip}{0pt}
\caption{Schematic diagrams of the kinetic energy flux $\Pi_u(k)$ for the stably stratified system and convective system. (a) In stably stratified flows,  $\Pi_u(k)$ decreases with $k$  due to the negative energy supply rate $F_B(k)$. (b) In convective system, $F_B(k)>0$, hence  $\Pi_u(k)$ first increases for  $k < k_t$ where $F_B(k) > D(k)$, then $\Pi_u(k) \approx \mathrm{constant}$ for $k_t < k < k_d$ where $F_B(k) \approx D(k)$; $\Pi_u(k)$ decreases for $k>k_d$ where $F_B(k) < D(k)$. Reprinted with permission from Kumar {\em et al.}~\cite{Kumar:PRE2014}.}
\label{fig:sch_flux}
\end{figure}

\item For the stably stratified flows,  as argued by Bolgiano~\cite{Bolgiano:JGR1959} and Obukhov~\cite{Obukhov:DANS1959}, in the  $k_f < k< k_B$ wavenumber band,  buoyancy converts the kinetic energy of the flow to the potential energy, i.e., $F_B(k)  < 0$.  Hence, Eq.~(\ref{eq:Pi_k_Fk_Dk}) predicts that $\Pi_u(k)$ will decrease with $k$ in this wavenumber range, as shown in Fig.~\ref{fig:sch_flux}{(a)}. In the wavenumber range, $k_B < k < k_d$, buoyancy becomes weaker, hence $\Pi_u(k) \approx \mathrm{constant}$.  
\label{enu_2}

\item For RBC, buoyancy feeds  the kinetic energy, hence  $F_B(k)  > 0$.  Therefore we expect the KE flux $\Pi_u(k)$ to increase. Numerical simulation of Kumar {\em et al.}~\cite{Kumar:PRE2014} however show that the energy supplied by buoyancy is dissipated by the viscous force, i.e., $F_B(k) \approx D(k)$.  Hence $\Pi_u(k) \approx \mathrm{constant}$ in the inertial range, and we recover Kolmogorov's spectrum for RBC.  Note that L'vov~\cite{Lvov:PRL1991} argued that $F_B(k)<0$, which is not the case.
\label{enu_3}
\end{enumerate}

\subsubsection{Modeling and field theory:}

Researchers~\cite{Kraichnan:JFM1959,Yakhot:JSC1986,DeDominicis:PRA1979,Mccomb:JPA1985,Leslie:book,Mccomb:book:Turbulence} employed field-theoretic techniques to understand the physics of turbulent fluid. In field theory, the nonlinear terms of the equations are expanded perturbatively. Some of the popular field-theoretic techniques are direct interaction approximation (DIA)~\cite{Kraichnan:JFM1959,Leslie:book}, renormalization group analysis~\cite{Kraichnan:JFM1959,Yakhot:JSC1986,DeDominicis:PRA1979,Mccomb:JPA1985,Mccomb:book:Turbulence}, mean field approximation~\cite{Procaccia:PRL1989}, etc.  Field theory has been applied to buoyancy-driven flows as well.

As described in Sec.~\ref{sub:BO_GEN}, that Procaccia and Zeitak~\cite{Procaccia:PRL1989} employed mean field approximation to convective turbulence and obtained BO scaling. Rubinstein~\cite{Rubinstein:NASA1994} used Yakhot-Orszag's~\cite{Yakhot:JSC1986} renormalization group procedure and proposed an isotropic model for convective turbulence. His results are consistent with that of  Procaccia and Zeitak~\cite{Procaccia:PRL1989}. Recently, using self-consistent field theory, Bhattacharjee~\cite{Bhattacharjee:JSP2015} obtained $E_u(k) \sim k^{-13/3}$ for RBC in the infinite Prandtl number limit.  Bhattacharjee~\cite{Bhattacharjee:PLA2015} used  the global energy balance for the stratified fluid and argued that the BO scaling could be observed in stably stratified flow at high Richardson number. In addition, he also added the possibility of BO scaling for RBC in some range of Prandtl numbers. 

In the next section, we will present numerical results for the stably stratified turbulence and Rayleigh-B\'{e}nard convection.

\subsection{Numerical analysis of buoyancy-driven turbulence}

\subsubsection{Stably stratified turbulence:}
\label{sec:SSF}

Researchers simulated the stably-stratified turbulence  for the three regimes described in Subsection~\ref{subsec:BO_intro}.  First we discuss the results for a strong gravity that corresponds to $\mathrm{Ri} \gg 1$ or $\mathrm{Fr} \ll 1$.  Such configurations are observed in some regimes of plantery and stellar atmospheres.   Strong gravity makes such flows  quasi-2D with dual scaling, $k^{-3}$ and $k^{-5/3}$.  In this regime, Lindborg~\cite{Lindborg:JFM2006}, Brethouwer~\cite{Brethouwer:JFM2007}, and Bartello and Tobias~\cite{Bartello:JFM2013} showed that the spectra of the horizontal KE and PE follow $k_\perp^{-5/3}$ scaling, while the  energy spectrum of the vertical velocity follows $k_\parallel^{-3}$.  Vallgren {\em et al.}~\cite{Vallgren:PRL2011} included rotation in their simulation and obtained  KE spectra  $k^{-3}$ and $k^{-5/3}$ for two different wavenumber bands.

For weak stratification ($\mathrm{Ri} \ll 1$), Kumar {\em et al.}~\cite{Kumar:PRE2014} performed a 3D  stably-stratified turbulence simulation and reported Kolmogorov's spectrum for the kinetic energy as expected.  Kumar {\em et al.} also studied the moderate stratification regime and reported BO scaling, which will be described below.  In this paper we  focus on  the results for $\mathrm{Fr} \approx 1$ since they are recently observed.

\begin{figure}[htbp]
\begin{center}
\includegraphics[scale = 0.9]{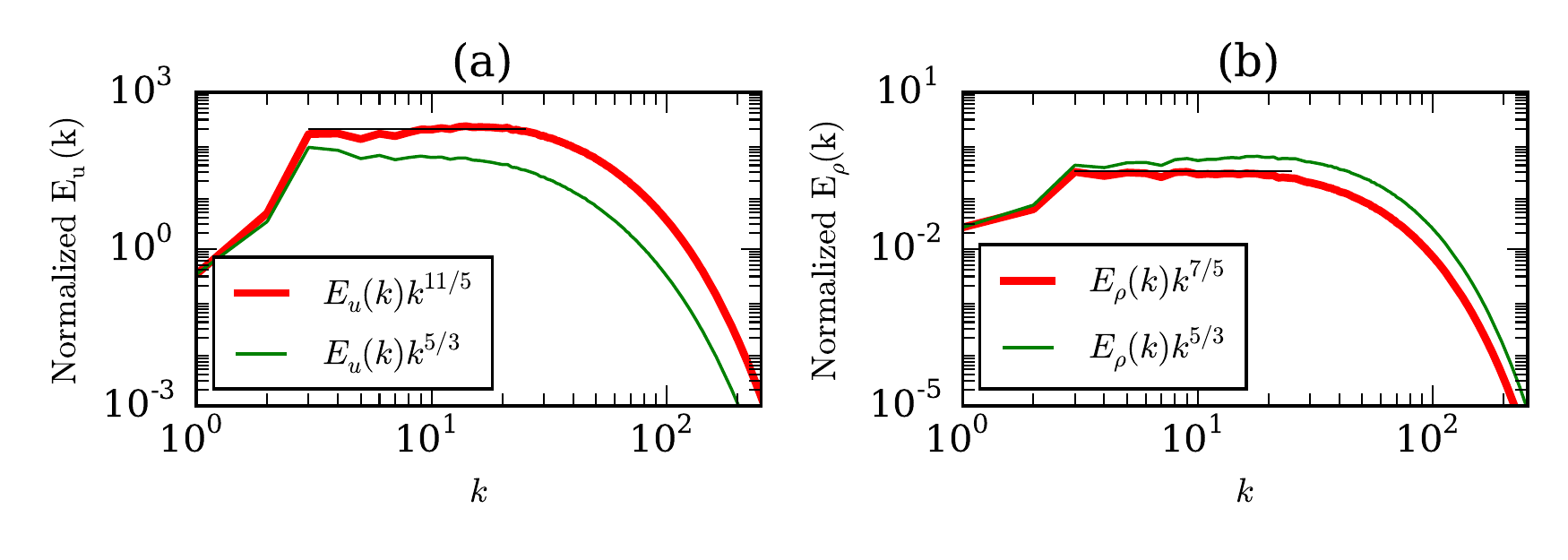}
\end{center}
\setlength{\abovecaptionskip}{0pt}
\caption{For the stably-stratified turbulence with $\mathrm{Pr}=1$, $\mathrm{Ra}=5 \times 10^3$, and $\mathrm{Fr} = 10$, plots of (a) normalized KE and (b) PE spectra for Bolgiano-Obukhov ({BO}) and Kolmogorov-Obukhov ({KO}) scaling.  {BO} scaling fits better with the data than KO scaling.  Reprinted with permission from Kumar {\em et al.}~\cite{Kumar:PRE2014}.}
\label{fig:spectra_0_01}
\end{figure}

Kumar {\em et al.}~\cite{Kumar:PRE2014} simulated  stably stratified flows in a cubical box of size $(2\pi)^3$ with  periodic boundary conditions at all the walls.  They forced the small wavenumber modes randomly to achieve a steady state. The parameters of their simulations are $\mathrm{Ra}=5 \times 10^3$ and $\mathrm{Pr}=1$ that yields $\mathrm{Ri}=0.01$ and $\mathrm{Fr}=10$. Figure \ref{fig:spectra_0_01}(a) exhibits the normalized KE spectra---$E_u(k)k^{11/5}$ for the {BO} scaling, and $E_u(k)k^{5/3}$ for the {KO} scaling.  The numerical data fits better with the {BO} scaling  for than the KO scaling, thus confirming the BO phenomenology  for the stably-stratified turbulence when $\mathrm{Fr} \approx 1$.  This is also verified by the PE spectrum as shown in  Fig.~\ref{fig:spectra_0_01}(b) in which $E_\rho(k)k^{7/5}$  provides a better fit to the data than $E_\rho(k)k^{5/3}$.

\begin{figure}[htbp]
\begin{center}
\includegraphics[scale = 1]{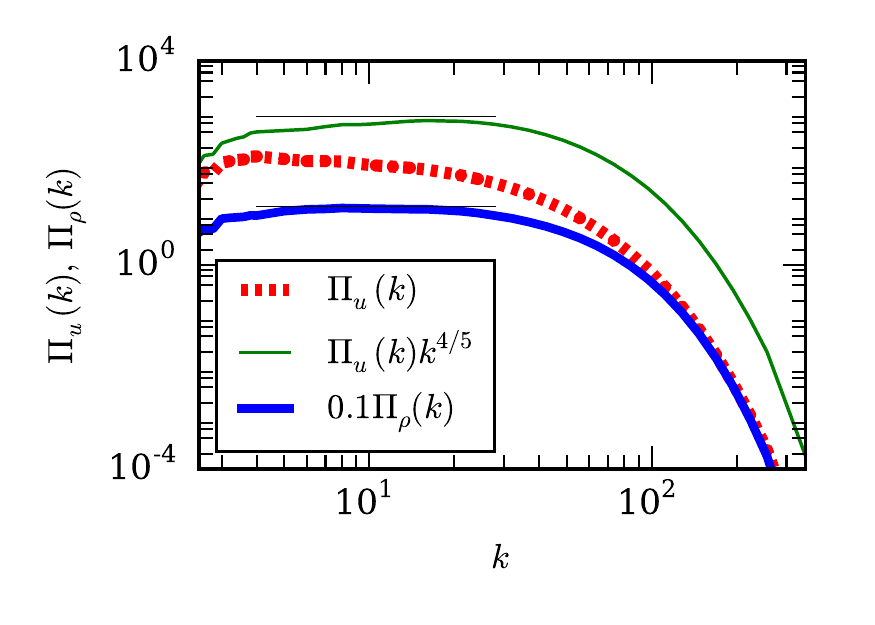}
\end{center}
\setlength{\abovecaptionskip}{0pt}
\caption{For  the stably-stratified turbulence with $\mathrm{Pr}=1$, $\mathrm{Ra}=5 \times 10^3$ and $\mathrm{Fr} = 10$  on $1024^3$ grid, plots of KE flux $\Pi_u(k)$, normalized KE flux $\Pi_u(k)k^{4/5}$, and potential energy flux $\Pi_{\rho}(k)$. The energy fluxes are also consistent with the BO phenomenology.  Reprinted with permission from Kumar {\em et al.}~\cite{Kumar:PRE2014}.}
\label{fig:flux_strat}
\end{figure}

\begin{figure}[htbp]
\begin{center}
\includegraphics[scale = 0.9]{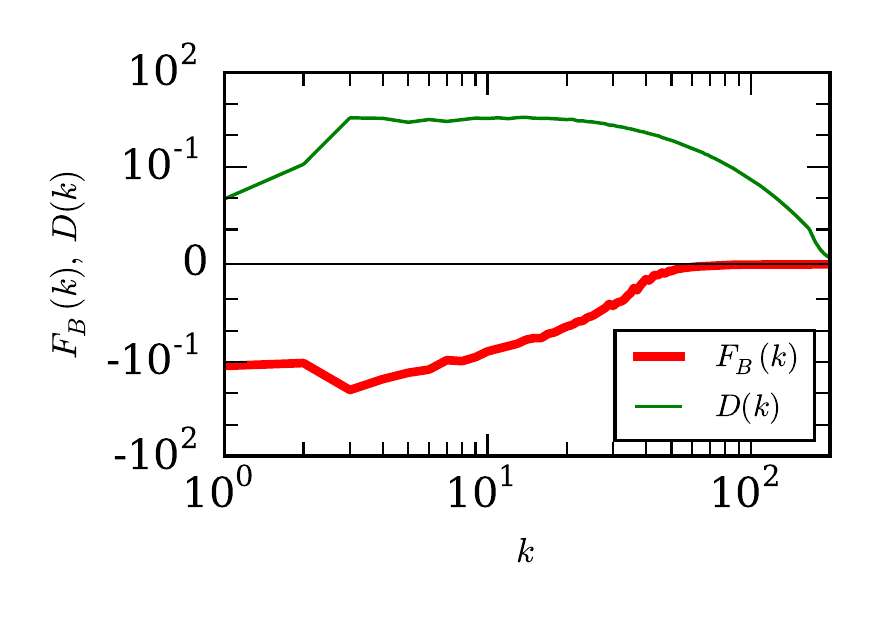}
\end{center}
\setlength{\abovecaptionskip}{0pt}
\caption{For  the stably-stratified turbulence with $\mathrm{Pr}=1$, $\mathrm{Ra}=5 \times 10^3$, and $\mathrm{Fr}=10$  on $1024^3$ grid, plots of the energy supply rate by buoyancy, $F_B(k)$, and the dissipation spectrum, $D(k)$.}
\label{fig:deriv_1}
\end{figure}

Further, Kumar {\em et al.}~\cite{Kumar:PRE2014} computed the KE and PE fluxes which are exhibited in Fig.~\ref{fig:flux_strat}.  They observed that $\Pi_u(k) > 0$ and it decreases with $k$ [Eq.~(\ref{eq:pi})], while the PE flux  $\Pi_\rho$ is a constant in the inertial range [Eq.~(\ref{eq:pi_theta})]; thus flux results are consistent with the {BO} predictions.    Kumar {\em et al.}~\cite{Kumar:PRE2014} also computed the energy supply rate by buoyancy, $F_B(k)$, and the viscous dissipation spectrum, $D(k)$, which are illustrated in  Fig.~\ref{fig:deriv_1}.  Note that $F_B(k) < 0$, as argued in BO phenomenology.    The Bolgiano wavenumber $k_B$  of Eq.~(\ref{eq:kb}) is approximately $8.5$, which is only 3 to 4 times smaller than $k_d$, wavenumber where the dissipation range starts.  Therefore   Kumar {\em et al.}~\cite{Kumar:PRE2014} did not observe a definitive crossover from $k^{-11/5}$ to $k^{-5/3}$ in their simulations. 

The aforementioned observations  demonstrate  applicability of the BO scaling for SST with a moderate stratification.

\subsubsection{Rayleigh-B\'{e}nard Convection:}
\label{subsec:RBC_Ek_num}

A large number of numerical simulations have been performed with an aim to identify which among the two, BO or KO, scaling is applicable to RBC.  Grossmann and Lohse~\cite{Grossmann:PRL1991} using simulation for $\mathrm{Pr}=1$ under Fourier-Weierstrass approximation and reported Kolmogorov's  scaling. Based on periodic boundary condition, Borue and Orszag~\cite{Borue:JSC1997} and \v{S}kandera {\em et al.}~\cite{Skandera:HPCISEG2SBH2009} reported KO scaling for the velocity and temperature fields. Kerr~\cite{Kerr:JFM1996}  reported the horizontal spectrum as a function of horizontal wavenumber  and observed Kolmogorov's spectrum. Verzicco and Camussi~\cite{Verzicco:JFM2003b}, and Camussi and Verzicco~\cite{Camussi:EJMB2004} showed BO scaling using the frequency spectrum of real space probe data. Kaczorowski and Xia~\cite{Kaczorowski:JFM2013} reported KO scaling for the longitudinal velocity structure functions, but BO scaling for the temperature structure functions in the centre of a cubical cell.  Kumar {\em et al.}~\cite{Kumar:PRE2014} computed $E_u(k)$ and $\Pi_u(k)$, and showed Kolmogorov-like behaviour for RBC, i.e., $E_u(k) \sim k^{-5/3}$ and $\Pi_u(k) \sim \mathrm{const}$.  In this paper we present the above quantities for $4096^3$ resolution and very high $\mathrm{Ra}$  that unambiguously demonstrates KO scaling for RBC.  We also report  the shell-to-shell energy transfers and the ring spectrum for RBC that show close resemblance with the hydrodynamic turbulence.

\begin{figure}[htbp]
\begin{center}
\includegraphics[scale = 0.9]{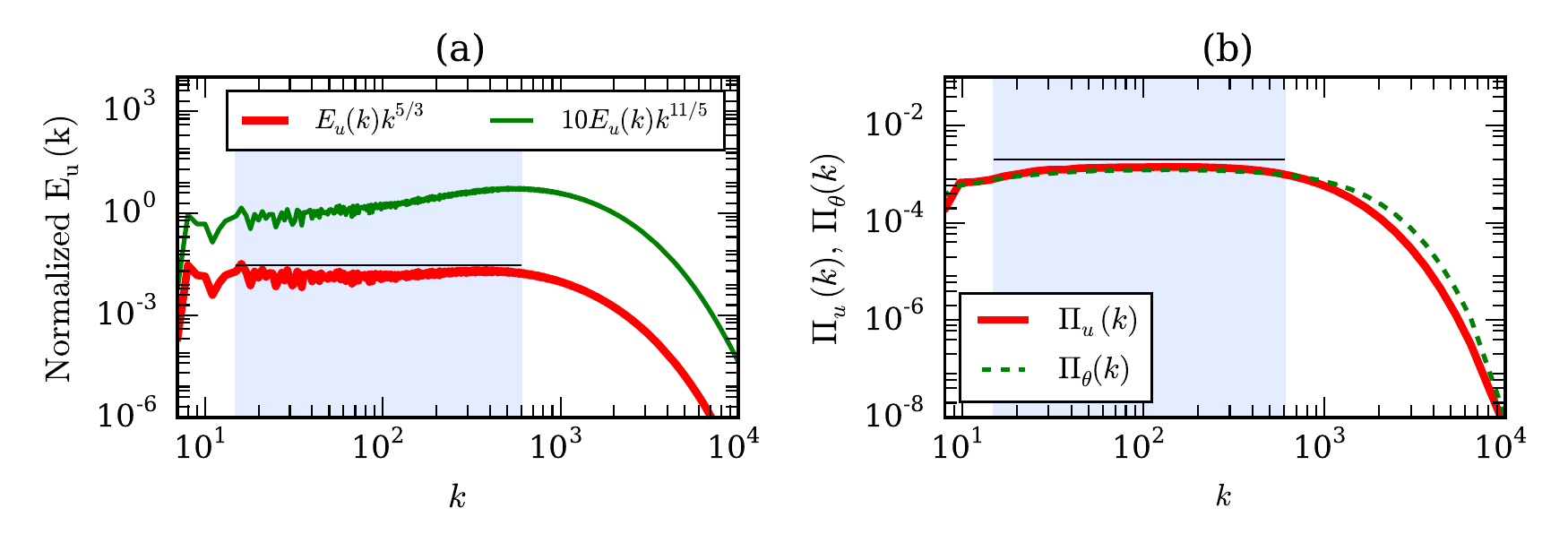}
\end{center}
\setlength{\abovecaptionskip}{0pt}
\caption{For the RBC simulation with $\mathrm{Pr}=1$ and $\mathrm{Ra}=1.1 \times 10^{11}$  on $4096^3$ grid: (a) plots of normalized KE spectra for Bolgiano-Obukhov ({BO}) and Kolmogorov-Obukhov ({KO}) scaling; {KO} scaling fits better with the data than {BO} scaling. (b) KE flux $\Pi_u(k)$ and entropy flux $\Pi_\theta(k)$. The shaded region exhibits the inertial range.}
\label{fig:ke_spectrum_RBC}
\end{figure}

We performed RBC simulations  in a unit box with $4096^3$ grid for  $\mathrm{Pr}=1$ and $\mathrm{Ra} = 1.1 \times 10^{11}$.  For the velocity field, we employed the free-slip boundary condition at the top and bottom plates, and periodic boundary condition at the side walls.  The temperature field satisfies conducting boundary condition at the top and bottom plates, and the periodic boundary condition at the side walls.  We computed the spectra and fluxes of the KE and the entropy ($\theta^2/2$) using the steady state data. Figure~\ref{fig:ke_spectrum_RBC}{(a)} exhibits the KE spectra normalized with $k^{11/5}$ and $k^{5/3}$. The plots indicate that in the wavenumber band $15 < k < 600$ (inertial range), the shaded region of the figure, the {KO} scaling fits better than the {BO} scaling.
 
 \begin{figure}[htbp]
\begin{center}
\includegraphics[scale = 0.9]{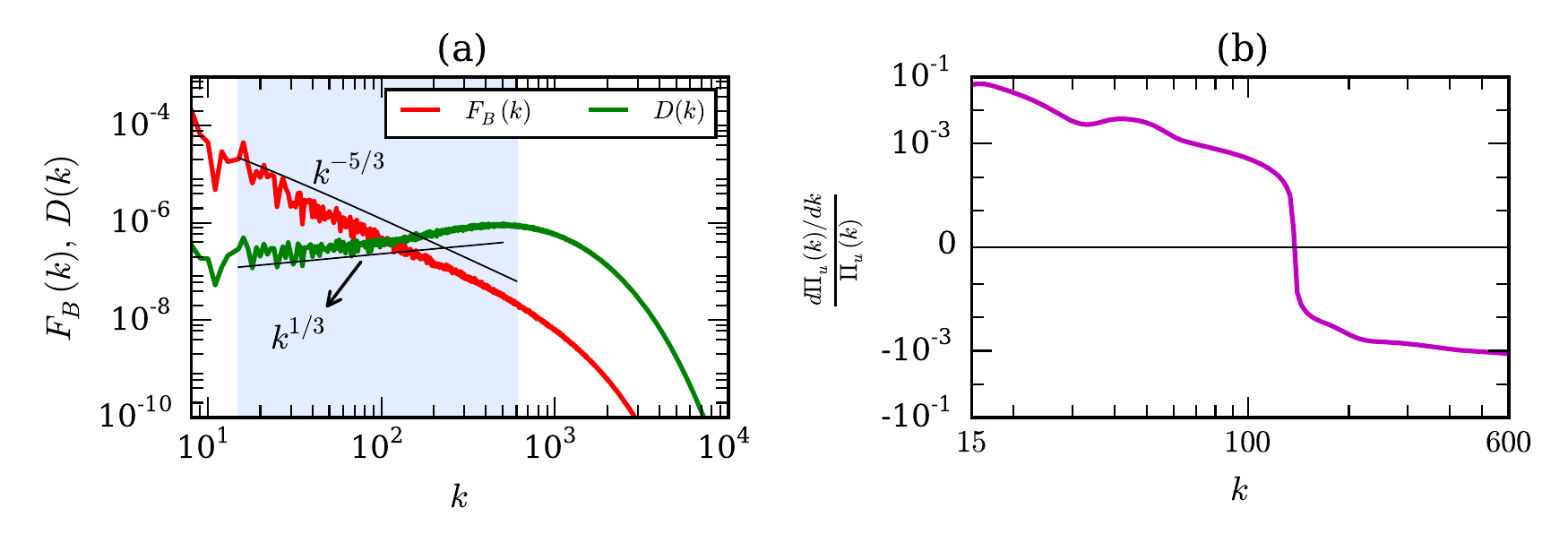}
\end{center}
\setlength{\abovecaptionskip}{0pt}
\caption{For the RBC simulation with $\mathrm{Pr}=1$ and $\mathrm{Ra} = 1.1 \times 10^{11}$: (a) plots of $F_B(k)$ and $D(k)$. (b) plots of $[d\Pi_u(k)/dk]/{\Pi_u(k)}$ in the inertial range $15<k<600$. }
\label{fig:force_RBC}
\end{figure}

 We exhibit the KE and entropy fluxes in Fig~\ref{fig:ke_spectrum_RBC}{(b)}.  We observe that the kinetic energy flux $\Pi_u(k)$ remains constant in the inertial range, a band where $E_u(k) \sim k^{-5/3}$.     Thus we claim that the convective turbulence exhibits Kolmogorov's power law in the inertial range.     We also computed $F_B(k)$, $\Pi_u(k)$, and $d\Pi_u(k)/dk$ as further tests.  According to Fig.~\ref{fig:force_RBC}(a) $F_B(k) > 0$ in the inertial range,  consistent with the discussion of Sec.~\ref{sec:pheno} and Fig.~\ref{fig:sch_flux}{(b)}, and it approximately balances $D(k)$.      Therefore, $d\Pi_u(k)/dk \approx 0$ or $\Pi_u(k) \approx \mathrm{constant}$ [see Eq.~(\ref{eq:dPik_dk})]. The constancy of $\Pi_u(k)$ yields $E_u(k) \sim k^{-5/3}$, consistent with the energy spectrum plots of Fig.~\ref{fig:ke_spectrum_RBC}(a).  Fig.~\ref{fig:force_RBC}(b) shows that $[d\Pi_u(k)/dk]/{\Pi_u(k)} \ll 1$ in the inertial range consistent with the constant $\Pi_u(k)$.  Interestingly,  $D(k) = 2 \nu k^2 E_u(k) \sim k^{1/3}$, consistent with $E_u(k) \sim k^{-5/3}$.  Also, $F_B(k)  \sim k^{-5/3}$.  In addition, the entropy flux $\Pi_\theta(k)$ is  constant, and  $\Pi_u(k) \approx \Pi_\theta(k)$ in dimensionless units.

\begin{figure}[htbp]
\begin{center}
\includegraphics[scale = 0.9]{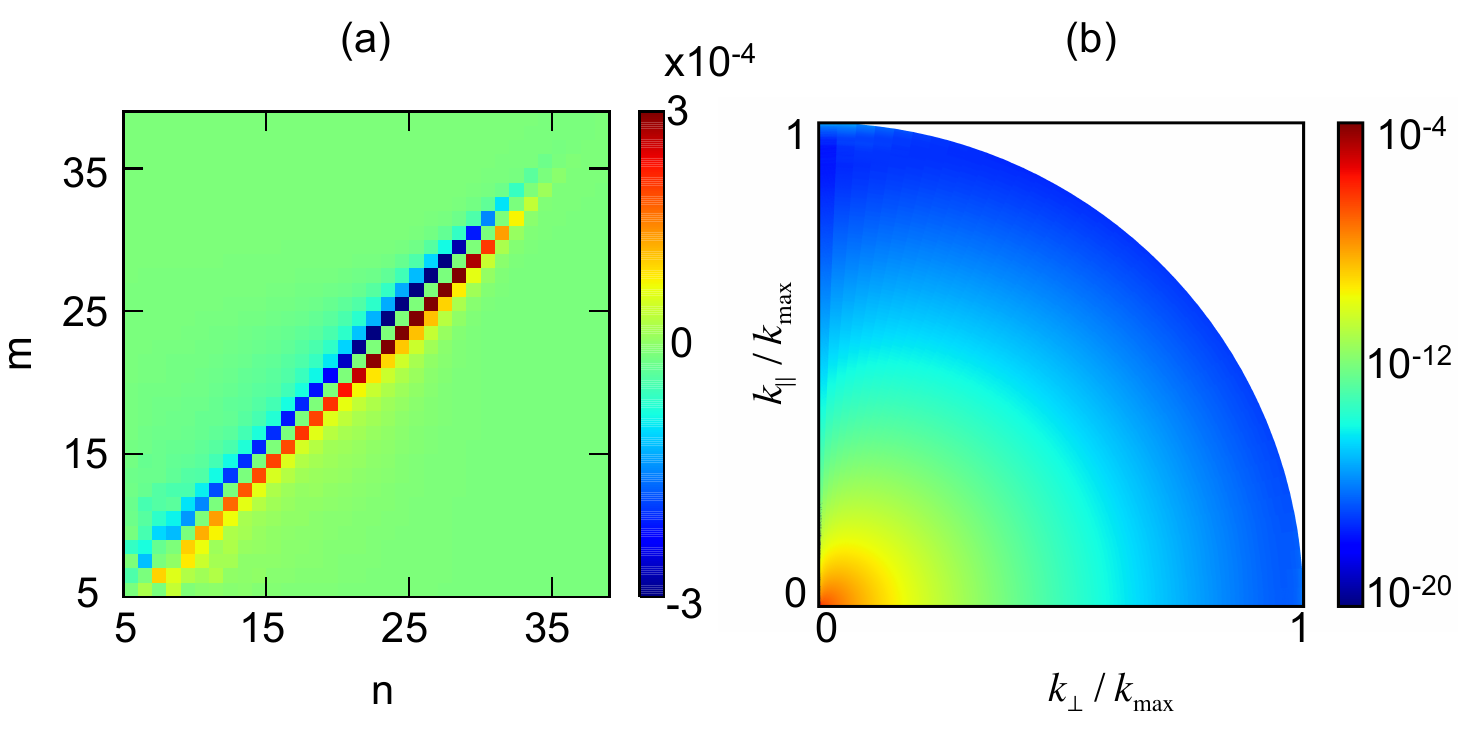}
\end{center}
\setlength{\abovecaptionskip}{0pt}
\caption{For the RBC simulation  for $\mathrm{Pr} = 1$ and $\mathrm{Ra} = 1.1 \times10^{11}$: (a) Plot of the shell-to-shell energy transfers $T^{m}_{n}$ of Eq.~(\ref{eq:S2S}), where $m,n$ represent the giver and receiver shell indices respectively.  (b)  Plot of the ring spectrum $E(k,\beta)$ demonstrates near isotropy in the Fourier space.}
\label{fig:RBC_shell_ring}
\end{figure}

We also compute the shell-to-shell energy transfers [Eq.~(\ref{eq:S2S})] using the  steady-state  data of our simulation.  We divide the Fourier space into $40$ concentric shells; the inner and outer radii of the $n$th shell are $k_{n-1}$ and $k_{n}$ respectively with  $k_{n} = \{0, 2, 4, 8, 8\times 2^{s(n-3)},..., 6432\}$, where $s=(1/35)\log_2(804)$.  The radii of the inertial-range shells are binned logarithmically due to the power law physics of RBC in the inertial range.  In Fig.~\ref{fig:RBC_shell_ring}(a) we exhibit the shell-to-shell energy transfers with the indices of the $x,y$ axes representing the receiver and giver shells respectively.  The plot indicates that  $m$th shell gives energy to $(m+1)$th shell, and it receives energy from the $(m-1)$th shell.  Thus the energy transfer in RBC is local and forward, very similar to hydrodynamic turbulence.  This result is consistent with the energy spectrum and flux studies described earlier.

Convective flows are expected to be anisotropic due to buoyancy; hence it is important to quantify anisotropy using quantities that are dependent on the polar angle, the angle between $\hat{z}$ and ${\bf k}$. For the same, we divide a wavenumber shell into rings~\cite{Nath:arxiv2016}. The energy contents of the rings are called {\em ring spectrum} $E(k,\beta)$, where $\beta$ represents the sector index for the polar angles (for details see Nath {\em et al.}~\cite{Nath:arxiv2016}). The ring spectrum $E(k,\beta)$, depicted in Fig.~\ref{fig:RBC_shell_ring}(b), shows that the flow is nearly isotropic, again similar to hydrodynamic turbulence. These results clearly demonstrate that the turbulent convection for $\mathrm{Pr}=1$ has a very similar behavior as hydrodynamic turbulence. 

\begin{figure}[htbp]
\begin{center}
\includegraphics[scale = 0.9]{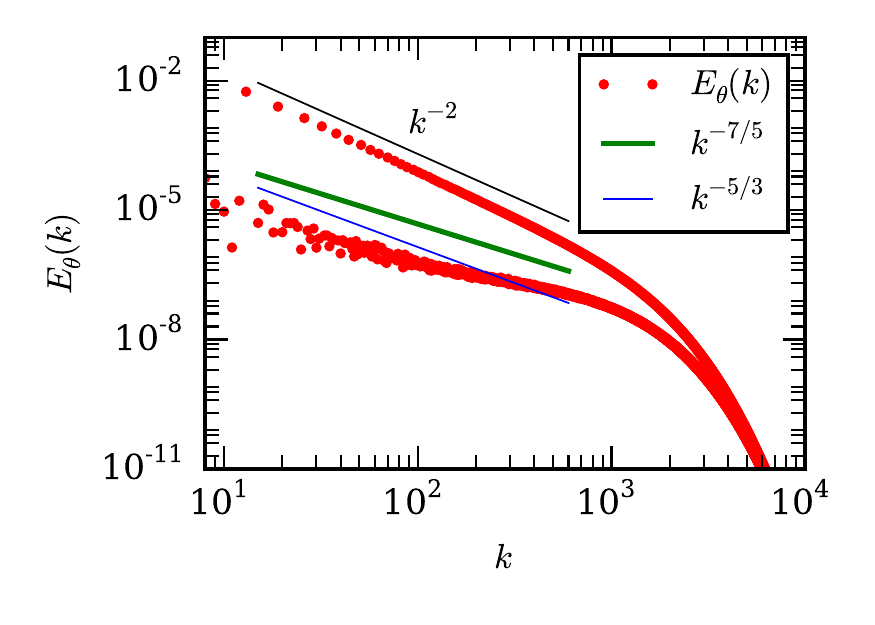}
\end{center}
\setlength{\abovecaptionskip}{0pt}
\caption{For RBC simulation with $\mathrm{Pr}=1$ and $\mathrm{Ra}=1.1 \times 10^{11}$, plot of the entropy spectrum that exhibits dual branches. The upper branch matches with $k^{-2}$ quite well, while the lower part is fluctuating.}
\label{fig:entropy_RBC}
\end{figure}

The temperature fluctuation however exhibit a unique behaviour.  As illustrated in Figure~\ref{fig:entropy_RBC}, we observe  dual branches for the entropy spectrum ($E_{\theta}(k)$).  The upper branch varies as $k^{-2}$ because $\theta(0,0,k_z) \approx -1/(\pi k)$, as discussed in Sec.~\ref{subsec:T(z)}.  The lower branch shows neither  KO ($k^{-5/3}$) nor BO ($k^{-7/5}$) spectrum. Note that both the branches of entropy spectrum generate a constant entropy flux $\Pi_{\theta}(k)$ (see Fig.~\ref{fig:ke_spectrum_RBC}(b)), and the modes $\theta(0,0,k_z)$ also participate in energy transfers.


\subsection{Experimental results}
For stably-stratified flows, there are not many laboratory experiments to verify BO phenomenology.  However, scientists have measured the KE spectrum of the Earth's atmosphere and relate it to the theoretical predictions.  Most notably Gage and Nastrom~\cite{Gage:JAS1986} observed a combination of $k^{-3}$ and $k^{-5/3}$ energy spectra.   Some researchers attribute the $k^{-3}$ spectrum at lower wavenumbers to the two-dimensionalization of the flow, while $k^{-5/3}$ spectrum at larger wavenumbers to the forward cascade of kinetic energy; yet these issues are still unresolved.  These features  are expected to arise for $\mathrm{Fr} \ll 1$.  

There are a significant number of laboratory experiments on RBC, with some favouring the BO scaling~\cite{Chilla:NC1993,Zhou:PRL2001}, while some others in support of  the KO scaling~\cite{Cioni:EPL1995}.  In most convective experiments, the velocity field, $u_z({\bf r}, t)$, and/or the temperature field, $T({\bf r}, t)$, are probed near the lateral walls of the container. For such experiments, the Taylor's hypothesis~\cite{Taylor:PRSLA1938,Shang:PRE2001,Kumar:2016a} is invoked to relate the frequency power spectrum $E(f)$ to the one-dimensional wavenumber spectrum $E(k)$; this connection is under debate due to the absence of any constant mean velocity field.   Researchers \cite{Kunnen:PRE2008,Sun:PRL2006,Zhou:JFM2011,Zhou:JFM2008} employ 2D particle image velocimetry (PIV) for high-resolution visualization and  computation of an approximate energy spectrum under the assumption of homogeneity and isotropy, which is not  strictly valid in convection~\cite{Nath:arxiv2016}.  In summary, on the experimental front, there is no convergence on which of the two scaling, BO of KO, is valid.  For details refer to the review papers~\cite{Ahlers:RMP2009,Lohse:ARFM2010}.

\subsection{Turbulence in thermal boundary layer and in two dimensions}
\label{subsec:Ek2D}

\begin{figure}[htbp]
\begin{center}
\includegraphics[scale = 0.7]{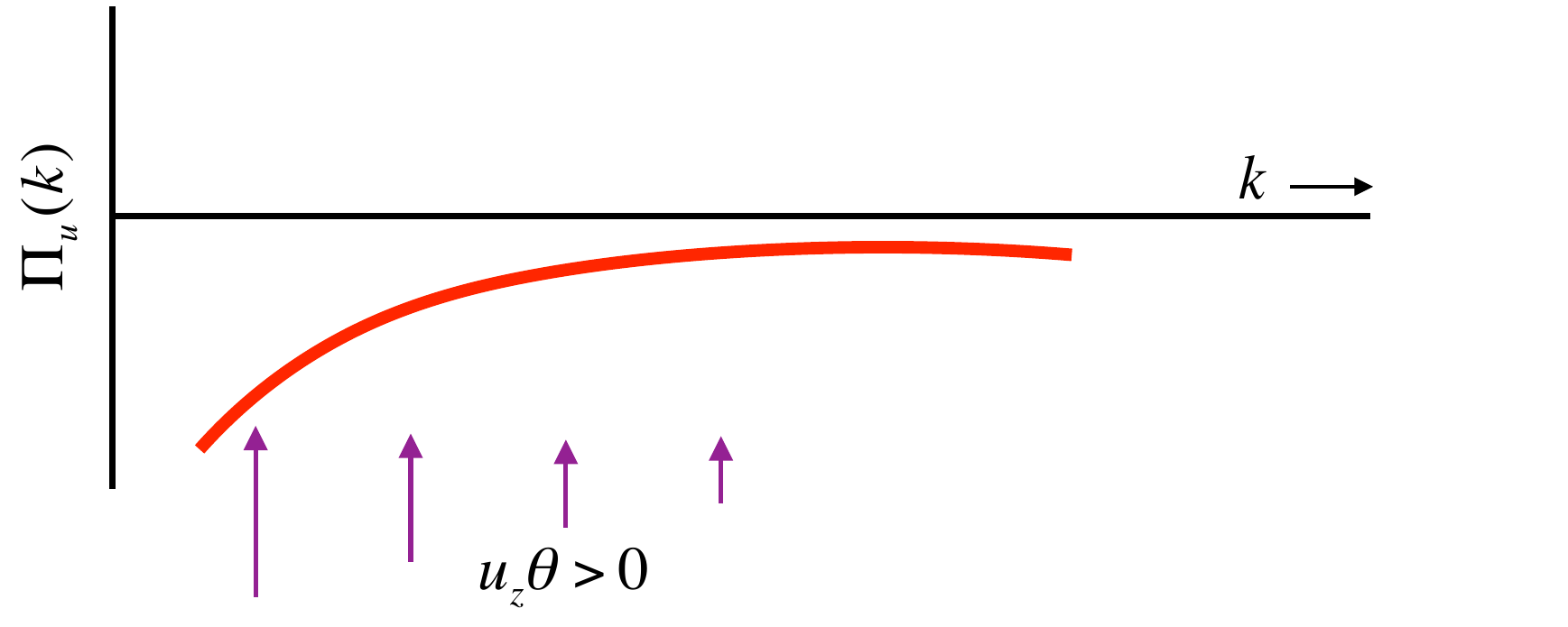}
\end{center}
\setlength{\abovecaptionskip}{0pt}
\caption{A possible schematic diagram of the kinetic energy flux $\Pi_u(k)$ for two-dimensional RBC.}
\label{fig:sch_2D_RBC}
\end{figure}

A burning question is whether KO or BO scaling are applicable to the boundary layers of RBC.   The flux arguments of Sec.~\ref{sec:pheno} provide some insights into the dynamics of boundary layers. 
Here, typically $u_z \ll u_\perp$, hence the flow is quasi-2D, and we expect an inverse cascade of KE.  Using $\Pi_u(k) < 0$, $F_B(k) > 0$, and $d\Pi_u(k)/dk \approx F_B(k)$, we may argue that  $\Pi_u(k)$ may increase with $k$ as shown in Fig.~\ref{fig:sch_2D_RBC}. An application of scaling arguments of  Sec.~\ref{subsec:BO_intro} may yield  spectra and  fluxes according to Eqs.~(\ref{eq:Eu}-\ref{eq:pi_theta}), i.e., Bolgiano-Obukhov scaling for $k <k_B$.  For $k > k_B$, the KE spectrum may exhibit a mixture of $k^{-5/3}$ (regime of inverse cascade of energy) and $k^{-3}$ (regime of forward cascade of enstrophy) depending on where the effective forcing band lies in relation to $k_B$.  Thus, in the boundary layer, RBC may exhibit BO scaling, and it needs to be investigated carefully using numerical simulations and experiments.    

The aforementioned scaling arguments may work for 2D RBC ($xz$ plane in which the buoyancy is along the $z$ direction), as well as in quasi 2D RBC (when $L_x \gg L_y$). Toh and Suzuki~\cite{Toh:PRL1994} simulated 2D RBC and reported $E_u(k) \sim k^{-11/5}$  and $\Pi_u(k) \sim -k^{-4/5}$ in line with the above arguments. Calzavarini {\em et al}.~\cite{Calzavarini:PRE2002} also reported similar results in their structure function computations.

\subsection{Turbulence in Rayleigh-Taylor instability (RTI)}
\label{subsec:RTI}
Chertkov~\cite{Chertkov:PRL2003} proposed that a fully-developed 3D RTI will exhibit Kolmogorov's spectrum due to the Rayleigh-Taylor pumping at large scales.  Boffetta {\em et al.}~\cite{Boffetta:PRE2009} observed this behaviour in their numerical simulations.   Chertkov~\cite{Chertkov:PRL2003} however does not take into account the buoyancy at all scales (see Sec.~\ref{sec:pheno}).   In a quasi-2D box ($L_y \ll L_x$), Boffetta {\em et al.}~\cite{Boffetta:JFM2011} show coexistence of BO and KO scaling ($k^{-11/5}$ and $k^{-5/3}$), consistent with the arguments of Sec.~\ref{subsec:Ek2D}.

\subsection{Turbulence in miscellaneous systems}
Scientists have studied  spectra of the velocity and the scalar field in other buoyancy-driven systems.  Pawar and Arakeri~\cite{Pawar:PF2016} performed experiment on the vertical tube described in Sec.~\ref{subsec:Arakeri}.  They observed that the velocity field exhibits $k^{-5/3}$ spectrum, while the scalar spectrum is closer to $k^{-7/5}$.   

Prakash {\em et al.}~\cite{Prakash:JFM2016} studied the energy spectrum of the bubbly turbulence using an experiment.  For the velocity field, they reported $k^{-5/3}$ energy spectrum for $k < 1/b$, and $k^{-3}$ for $k > 1/b$ where $b$ is the bubble size.  They argue that that the large and intermediate scales exhibit $k^{-5/3}$ spectrum due to the standard Kolmgorov's argument.  For $k > 1/b$, Prakash {\em et al.}~\cite{Prakash:JFM2016} explained the $k^{-3}$ energy spectrum by invoking equipartition between the energy dissipation and energy feed by the buoyancy.  For this system it may be interesting to investigate the energy spectrum using the flux arguments. 

The  turbulent Taylor-Couette flow~\cite{Grossmann:ARFM2016} may exhibit spectral behaviour similar to RBC since both the systems are unstable with similar energetics (see Secs.~\ref{sec:pheno} and \ref{subsec:RBC_Ek_num}).  We believe that the Non-Bouusinesq convective flows may also exhibit Kolmogorov-like spectrum for weak compressibility since here too the thermal plumes feed the kinetic energy, as in RBC.

\subsection{Turbulence in small and large Prandtl number RBC}
In Sec.~\ref{sec:pheno} we derived the spectra and fluxes of the velocity and temperature for RBC when $\mathrm{Pr} \sim 1$.  These arguments are not applicable to RBC with extreme Prandtl numbers.  However, we can easily deduce the spectrum for very small and very large  $\mathrm{Pr} $'s as follows.  These computations have been first reported in~\cite{Mishra:PRE2010} and \cite{Pandey:PRE2014} respectively.

In RBC with  zero or small Prandtl numbers, thermal diffusivity $\kappa \rightarrow 0$ that leads to $u_z({\bf k}) \sim  \theta({\bf k})/(\kappa k^2) $~\cite{Mishra:PRE2010}.  Hence, the buoyancy, which is proportional to $\theta({\bf k})$, is dominant at small wavenumbers.  Therefore, the assumption of the Kolmogorov's phenomenology that the forcing acts at large length scales is valid, and we expect the Kolmogorov's phenomenology for the hydrodynamic turbulence  to be applicable to RBC with $\mathrm{Pr} \rightarrow 0$.  Mishra and Verma~\cite{Mishra:PRE2010} verified the above phenomenology using numerical simulations.

In the limit of infinite Prandtl number ($\nu \rightarrow \infty$), the momentum equation is linear~\cite{Pandey:PRE2014}.  However if the P\'{e}clet number is large, the temperature equation is  nonlinear and it yields an approximate constant entropy flux.  Using scaling arguments, Pandey {\em et al.}~\cite{Pandey:PRE2014} derived that for infinite and large $\mathrm{Pr}$, $E_u(k) \sim k^{-13/3}$.  They also verified the above scaling using numerical simulations.

\subsection{Simulation of turbulent convection in a periodic box and shell model}

Borue and Orszag~\cite{Borue:JSC1997}, \v{S}kandera {\em et al.}~\cite{Skandera:HPCISEG2SBH2009}, Lohse and Toschi~\cite{Lohse:PRL2003}, and Calzavarani {\em et al.}~\cite{Calzavarini:PF2005} simulated turbulent thermal convection in a periodic box.  They simulated Eqs.~(\ref{eq:NS_RBC},\ref{eq:continuity_RBC}) under a gradient $d\bar{T}/dz$.  In the absence of boundary layers, the velocity and temperature fields exhibit $k^{-5/3}$ spectra~\cite{Borue:JSC1997,Skandera:HPCISEG2SBH2009}. In addition, the Nusselt number $\mathrm{Nu} \sim \mathrm{Ra}^{1/2}$~\cite{Lohse:PRL2003,Calzavarini:PF2005}, which is expected in the ultimate regime when the effects of boundary layers are negligible. Note that the temperature spectrum for the periodic box is very different from that with conducting walls that exhibit dual spectra. It is important to note that turbulent thermal convection in a periodic box is numerically unstable; the system exhibits steady behaviour for carefully chosen set of initial conditions.

Direct numerical simulation of turbulent systems is quite demanding due a large number of interacting Fourier modes.  Therefore, scientists often use shell models, which are based on much fewer number of modes.  Brandenburg\cite{Brandenburg:PRL1992}, Lozhkin and Frick~\cite{Lozhkin:FD1998}, Mingshun and Shida~\cite{Mingshun:PRE1997}, Ching and Cheng~\cite{Ching:PRE2008c}, and Kumar and Verma~\cite{Kumar:PRE2015,Kumar:IUTAM2015} constructed shell models for buoyancy-driven turbulence.  The advantage of the shell model of Kumar and Verma~\cite{Kumar:PRE2015} is that it describes both turbulent stably-stratified  and convective flows using a single set of equations. It also enables  flux computation of the kinetic energy and density.  Kumar and Verma~\cite{Kumar:PRE2015} showed that the results of the shell model are consistent with the DNS results described earlier.

\subsection{Concluding remarks on the energy spectrum}

We summarise the  important results of this section in the following.
\begin{enumerate}
\item A large body of works on RBC assume Bolgiano-Obuknov scaling.  The flux-based arguments described in Secs.~\ref{sec:pheno} and \ref{subsec:RBC_Ek_num} clearly demonstrate that in three dimensions, RBC exhibits Kolmogorov-like energy spectrum.  Given this, Bolgiano length is not meaningful for RBC.  

\item Turbulence in RBC  has significant similarities with hydrodynamic turbulence.  For example, the KE flux is nearly constant in the inertial range; the shell-to-shell energy transfer is local and forward; the ring spectrum exhibits a near isotropy in Fourier space.   The constant KE flux is due to the near cancellation between the KE supply by buoyancy and the viscous dissipation rate.

\item Under nearly isotropic conditions (when Froude number is of the order of unity), the stably-stratified turbulence exhibits Bolgiano-Obukhov scaling.

\item The temperature fluctuations  exhibits dual spectra, with the upper branch scaling as $k^{-2}$.  In Sec.~\ref{subsec:T(z)} we discussed the origin of $k^{-2}$ spectrum in terms of the structures of the boundary layers and the bulk.
\end{enumerate}

\section{Modelling of large-scale quantities of RBC}
\label{sec:large_scales}

In this section we quantify the large-scale quantities of RBC, namely the Nusselt and Reynolds numbers.  Many researchers have worked on this problem; for details and references,  refer  to the review articles \cite{Ahlers:RMP2009,Bodenschatz:ARFM2000,Lohse:ARFM2010,Siggia:ARFM1994}.  Despite complexities of the flow, RBC exhibits certain universal behaviour; in the turbulent limit, $\mathrm{Pe} \sim \sqrt{\mathrm{Ra Pr}}$, but in the viscous regime, $\mathrm{Pe} \sim \mathrm{Ra}^{3/5}$~\cite{Grossmann:JFM2000,Pandey:2016a}.  The Nusselt number however scale as $\mathrm{Nu} \sim \mathrm{Ra}^\beta$ with $\beta$ ranging from 0.27 to 0.33.   Researchers have attempted to explain the above behaviour.   For brevity, in this review we only discuss  the models of Grossmann and Lohse (GL)~\cite{Grossmann:JFM2000, Grossmann:PRL2001, Grossmann:PRE2002, Grossmann:PF2004, Grossmann:PF2011} and that of Pandey {\em et al.}~\cite{Pandey:2016a}.

\subsection{Grossmann-Lohse model}
Grossmann and Lohse (GL)~\cite{Grossmann:JFM2000, Grossmann:PRL2001, Grossmann:PRE2002, Grossmann:PF2004, Grossmann:PF2011,Stevens:JFM2013}    derived the formulas for $\mathrm{Nu(Ra, Pr)}$ and $\mathrm{Re(Ra,Pr)}$ by exploiting the fact that the global viscous dissipation rate, $\epsilon_u$, and thermal dissipation rate, $\epsilon_T$,  get contributions from the bulk and boundary layers, i.e.,
\begin{eqnarray}
\epsilon_u & = & \epsilon_{u, BL} + \epsilon_{u, bulk}, \\
\epsilon_T & = & \epsilon_{T, BL} + \epsilon_{T, bulk},
\end{eqnarray}
where $BL$ and $bulk$  denote the  boundary layer and the bulk respectively. They invoked the exact relations of Shraiman and Siggia~\cite{Shraiman:PRA1990} for the global viscous and thermal dissipation rates (see Eqs.~(\ref{eq:eps_u},\ref{eq:eps_theta})), and estimated the aforementioned contributions of the boundary layers and the bulk to $\epsilon_u $ and $\epsilon_T$ in various $\mathrm{Ra}$-$\mathrm{Pr}$ regimes.  For $\mathrm{Pr} \approx 1$ and very large $\mathrm{Ra}$ they used $\epsilon_{u,bulk} = U^3/d$ and $\epsilon_{T,bulk} = U\Delta^2/d$, but for extreme Prandtl numbers, these estimates get altered by the boundary layer widths. 

Using the above ideas, GL~\cite{Grossmann:JFM2000, Grossmann:PRL2001, Grossmann:PRE2002, Grossmann:PF2004, Grossmann:PF2011,Stevens:JFM2013} derived two coupled equations 
\begin{eqnarray}
\mathrm{(Nu-1)RaPr}^{-2} & = & c_1  \frac{\mathrm{Re}^2}{g(\sqrt{\mathrm{Re}_L/\mathrm{Re}})}  + c_2 \mathrm{Re}^3,  \\
\mathrm{Nu-1} & = & c_3 \sqrt{\mathrm{RePr}} \left \{ f \left[ \frac{2a\mathrm{Nu}}{\sqrt{\mathrm{Re}_L}} g \left ( \sqrt{\frac{\mathrm{Re}_L}{\mathrm{Re}}} \right) \right] \right \}^{1/2} \nonumber \\ 
& & + c_4 \mathrm{RePr} f \left[ \frac{2a\mathrm{Nu}}{\sqrt{\mathrm{Re}_L}} g  \left(  \sqrt{\frac{\mathrm{Re}_L}{\mathrm{Re}}}  \right) \right],
\end{eqnarray}
where $c_i$'s and $\mathrm{Re}_L$ are constants. Here functions $f$ and $g$ model the thermal BL~\cite{Stevens:JFM2013}. Using the above formulae, GL computed the Nusselt and Reynold numbers as a function of $\mathrm{Ra}$ and $\mathrm{Pr}$ that agree with presently available experimental and numerical simulation results quite well~\cite{Ahlers:RMP2009}.

\subsection{An alternate derivation of P\'{e}clet number}

\begin{figure}
\begin{center}
\includegraphics[scale=1]{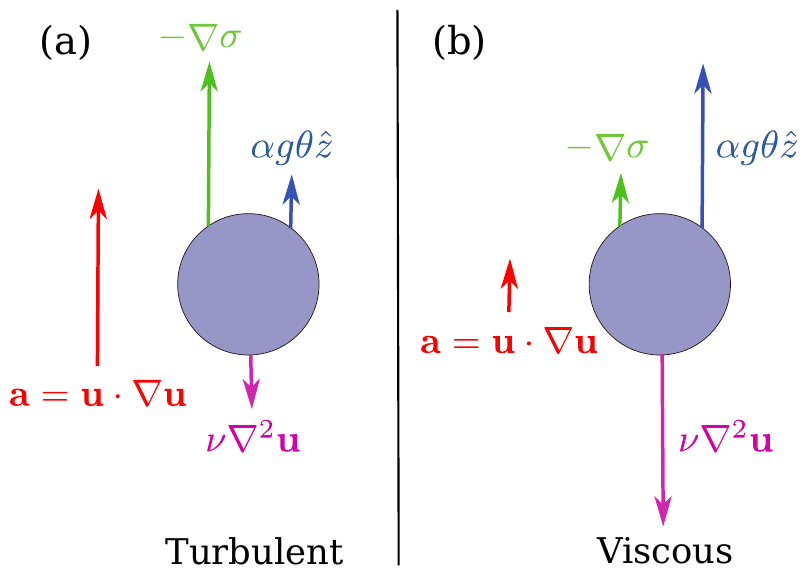}
\caption{The relative strengths of the forces acting on a fluid parcel. In the turbulent regime, the acceleration ${\bf u} \cdot \nabla {\bf u}$ is provided primarily by the pressure gradient.  In the viscous regime, the buoyancy and  the viscous force dominate the pressure gradient,  and they balance each other. Reprinted with permission from Pandey and Verma~\cite{Pandey:PF2016}.}
\label{fig:schematic_ns}
\end{center}
\end{figure}

Recently Pandey {\em et al.}~\cite{Pandey:2016a} and Pandey \& Verma~\cite{Pandey:PF2016} provided an alternate derivation of P\'{e}clet number.  Note that $\mathrm{Pe} = \mathrm{Re} \mathrm{Pr}$.  Pandey {\em et al.}~\cite{Pandey:2016a}  analysed the rms values of various terms of   the momentum equation, which  are exhibited in the schematic diagram of Fig.~\ref{fig:schematic_ns}.   Under  statistical steady state ($\langle \partial {\bf u}/\partial t \rangle \approx 0$),  Pandey {\em et al.} observed that in the turbulent regime, the acceleration ${\bf u} \cdot \nabla {\bf u}$ is primarily provided by the pressure gradient $-\nabla \sigma$, and  the buoyancy and viscous terms are relatively small.   The above features are consistent with similarities between the turbulence in RBC and hydrodynamics (see Sec.~\ref{subsec:RBC_Ek_num}).   However, in the viscous regime ($\mathrm{Re} \lessapprox 1$), $-\nabla \sigma$ is small, and the  buoyancy and viscous terms cancel each other resulting in a very small acceleration of the fluid.  

Dimensional analysis of the momentum equation yields
\begin{equation}
c_1 \frac{U^2}{d} =  c_2  \frac{U^2}{d}+  c_3 \alpha g \theta_{\mathrm{res}} - c_4 \nu \frac{U}{d^2}, \label{eq:U}
\end{equation}
where $c_i$'s are dimensionless coefficients defined as
\begin{eqnarray}
c_1 = \frac{|{\bf u \cdot \nabla u}|} {U^2/d};~~~c_2 = \frac{ |\nabla \sigma|_{\mathrm{res}} /\rho_0} {U^2/d};~~~c_3 = |\theta_{\mathrm{res}}/\Delta |;~~~c_4 =\frac{|\nabla^2 {\bf u}|} {U/d^2}.
\end{eqnarray}
Pandey {\em et al.}~\cite{Pandey:2016a} observed $c_i$'s to be functions of $\mathrm{Ra}$ and $\mathrm{Pr}$ that yields interesting and nontrivial scaling relations.   It is important to contrast this behaviour with free turbulence (without walls) where $c_i$'s are constants.   Multiplication of Eq.~(\ref{eq:U}) with $d^3/\kappa^2$ yields 
\begin{equation}
c_1 \mathrm{Pe}^2 = c_2 \mathrm{Pe}^2 + c_3 \mathrm{RaPr} - c_4 \mathrm{PePr},
\label{eq:Pe_eqn}
\end{equation}
where $\mathrm{Pe} = Ud/\kappa$ is the P\'{e}clet number.  The solution of the above equation is
\begin{equation}
\mathrm{Pe} = \frac{-c_4 \mathrm{Pr} + \sqrt{c_4^2 \mathrm{Pr}^2 + 4(c_1-c_2) c_3\mathrm{RaPr}}}{2 (c_1-c_2)}. \label{eq:Pe_analy}
\end{equation}	
 using which $\mathrm{Pe}$ can be computed as a function of $\mathrm{Ra}$ and $\mathrm{Pr}$.

In the turbulent regime, the viscous term of Eq.~(\ref{eq:Pe_eqn}) can be ignored, hence 
\begin{equation}
\mathrm{Pe} \approx \sqrt{\frac{c_3}{|c_1-c_2|} \mathrm{RaPr}}. 
\label{eq:Pe_turb}
\end{equation}	
This limit is applicable when 
\begin{equation}
c_4^2 \mathrm{Pr}^2 \ll  4 |c_1-c_2| c_3\mathrm{RaPr}.
\label{eq:Ra_cond_turb}
\end{equation}	
The scaling for the  viscous regime is obtained by equating the buoyancy and viscous terms of the momentum equation that yields
\begin{equation}
\mathrm{Pe} \approx \frac{c_3}{c_4} \mathrm{Ra}.
\label{eq:Pe_visc}
\end{equation}

\begin{figure}
\begin{center}
\includegraphics[scale=1.2]{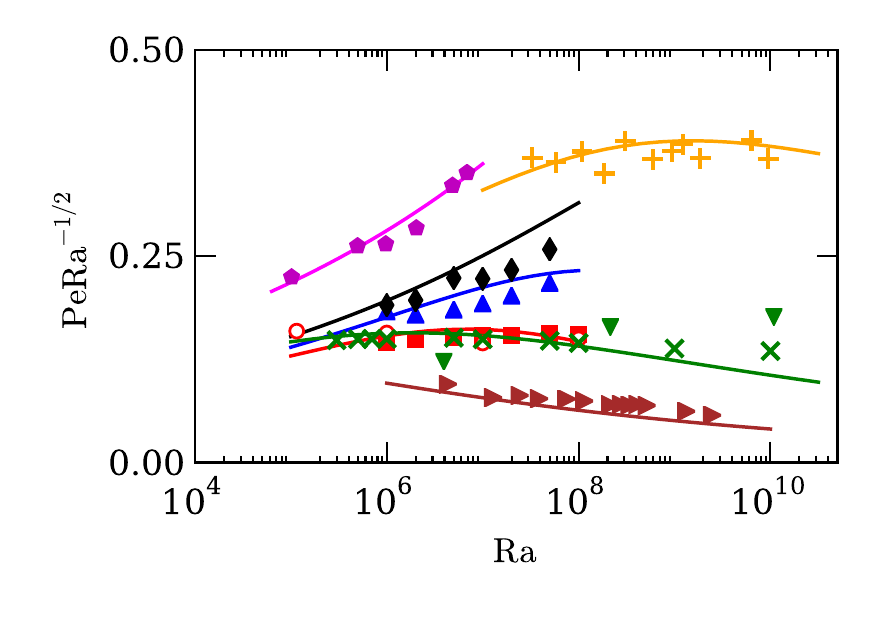}
\caption{The normalized P\'{e}clet number ($\mathrm{PeRa^{-1/2}}$) vs. $\mathrm{Ra}$ for  numerical data of Pandey {\em et al.}~\cite{Pandey:2016a} for $\mathrm{Pr} = 1$ (red squares), $\mathrm{Pr} = 6.8$ (blue triangles), and $\mathrm{Pr} = 10^2$ (black diamonds); numerical data of Silano \textit{et al.}~\cite{Silano:JFM2010} (magenta pentagons, $\mathrm{Pr} = 10^3$), Reeuwijk \textit{et al.}~\cite{Reeuwijk:PRE2008} (red circles, $\mathrm{Pr} = 1$), Scheel and Schumacher~\cite{Scheel:JFM2014} (green crosses, $\mathrm{Pr} = 0.7$); and the experimental data of Xin and Xia~\cite{Xin:PRE1997} (orange pluses, $\mathrm{Pr} \approx 6.8$), Cioni \textit{et al.}~\cite{Cioni:JFM1997} (brown right triangles, $\mathrm{Pr} \approx 0.022$), and Niemela \textit{et al.}~\cite{Niemela:JFM2001} ($\mathrm{Pr} \approx 0.7$, green down-triangles). The continuous curves represent $\mathrm{Pe}$ computed using  Eq.~(\ref{eq:Pe_analy}).  The predictions of Eq.~(\ref{eq:Pe_analy}) for $\mathrm{Pr} = 0.022$ and 6.8 have been multiplied with 2.5 and 1.2, respectively, to fit the experimental results from Cioni \textit{et al.}~\cite{Cioni:JFM1997} and Xin and Xia~\cite{Xin:PRE1997}. Reprinted with permission from Pandey and Verma~\cite{Pandey:PF2016}. }
\label{fig:pe_ns}
\end{center}
\end{figure}

Pandey {\em et al.}~\cite{Pandey:2016a}  computed $c_i$'s using the RBC simulation data for $\mathrm{Pr = 1,6.8, 10^2, 10^3}$ and $\mathrm{Ra}$ from $10^6$ to $5 \times 10^8$.  These simulations were performed for no-slip boundary condition at all the walls using a finite volume solver {\sc OpenFoam}~\cite{OpenFOAM}.  They reported the following functional form for $c_i$'s 
\begin{eqnarray}
c_1 & = & 1.5 \mathrm{Ra}^{0.10} \mathrm{Pr}^{-0.06}, \label{eq:c1_ns} \\
c_2 & = & 1.6 \mathrm{Ra}^{0.09} \mathrm{Pr}^{-0.08}, \label{eq:c2_ns} \\
c_3 & = & 0.75 \mathrm{Ra}^{-0.15} \mathrm{Pr}^{-0.05}, \label{eq:c3_ns} \\
c_4 & = & 20 \mathrm{Ra}^{0.24} \mathrm{Pr}^{-0.08}. \label{eq:c4_ns}
\end{eqnarray}
The errors in the above exponents are $\lessapprox 0.01$, except for the $\mathrm{Ra}$ exponent of  $c_4$ that has error of the order of 0.10. In Fig.~\ref{fig:pe_ns}, we plot the normalized P\'{e}clet number, $\mathrm{PeRa^{-1/2}}$ for $\mathrm{Pr = 1, 6.8, 10^2}$ and compare them with the predictions using Eq.~(\ref{eq:Pe_analy}).  The figure also exhibits  $\mathrm{Pe}$ from other simulations and experiments.  The plots reveal that the predictions of Pandey {\em et al.}~\cite{Pandey:2016a}  [Eq.~(\ref{eq:Pe_analy})] match with the numerical and experimental results quite well.

Using the above $c_i$'s and Eq.~(\ref{eq:Ra_cond_turb}), we find that $\mathrm{Ra \gg 10^6 Pr}$ belongs to the turbulent regime, whereas $\mathrm{Ra \ll 10^6 Pr}$ belongs to the viscous regime.   In the viscous regime
\begin{equation}
\mathrm{Pe} = \frac{c_3}{c_4} \mathrm{Ra} \approx  0.038 \mathrm{Ra}^{0.60}, \label{eq:Pe_visc_reg}
\end{equation}	
which is independent of $\mathrm{Pr}$, consistent with the results of Silano \textit{et al.}~\cite{Silano:JFM2010}, Horn \textit{et al.}~\cite{Horn:JFM2013},  and Pandey \textit{et al.}~\cite{Pandey:PRE2014}. For the turbulent regime, Eq.~(\ref{eq:Pe_turb}) yields  
\begin{equation}
\mathrm{Pe} = \sqrt{\frac{c_3}{|c_1-c_2|}} \sqrt{\mathrm{RaPr}} \approx \sqrt{\mathrm{7.5 Pr}} \mathrm{Ra}^{0.38}.
\label{eq:Pe_turb_0.38}
\end{equation}	
For mercury ($\mathrm{Pr} \approx 0.025$) as an experimental fluid, Cioni \textit{et al.}~\cite{Cioni:JFM1997} observed that $\mathrm{Re \sim Ra^{0.424}}$, which is close to the predicted exponent of 0.38 discussed above. The range of Rayleigh numbers in the experiment of Cioni \textit{et al.}~\cite{Cioni:JFM1997} is $5 \times 10^6 \le \mathrm{Ra} \le 5 \times 10^9$ that is consistent with   the turbulent regime estimated above ($\mathrm{Ra \gg 10^6 Pr}$).  The aforementioned results are in general agreement with those of Grossmann and Lohse~\cite{Grossmann:JFM2000, Grossmann:PRL2001, Grossmann:PRE2002, Grossmann:PF2004, Grossmann:PF2011}.

\subsection{Scaling of Nusselt number and dissipation rates}

The Nusselt number, a measure of the convective heat transport~\cite{Ahlers:RMP2009, Chilla:EPJE2012, Xia:TAML2013}, is quantified as
\begin{equation}
\mathrm{Nu} = \frac{\kappa \Delta/d + \langle u_z \theta_{\mathrm{res}} \rangle_V}{\kappa \Delta/d} = 1 + \left\langle \frac{u_z d}{\kappa} \frac{\theta_{\mathrm{res}}}{\Delta} \right\rangle_V = 1 + C_{u\theta_\mathrm{res}} \langle u_z^{'2} \rangle_V^{1/2} \langle \theta_\mathrm{res}^{'2} \rangle_V^{1/2}, \label{eq:Nu}
\end{equation}
where $\langle \rangle_V$ stands for a volume average,  $u'_z =  u_z d /\kappa$, $\theta'_{\mathrm{res}} = \theta_{\mathrm{res}}/\Delta$, and  $C_{u\theta_\mathrm{res}}$ is the normalized correlation function between the vertical velocity and the residual temperature fluctuation~\cite{Verma:PRE2012}:
\begin{equation}
C_{u\theta_\mathrm{res}} = \frac{\langle u'_z \theta'_\mathrm{res} \rangle_V} {\langle u_z^{'2} \rangle_V^{1/2} \langle \theta_\mathrm{res}^{'2} \rangle_V^{1/2}}.
\end{equation}
The deviation of the  exponent from 1/2 in the ultimate regime~\cite{Kraichnan:PF1962} is due to the nontrivial scaling of $C_{u\theta_\mathrm{res}}$, $u_z'$, and $\theta'_{\mathrm{res}}$. We observe that $C_{u\theta_\mathrm{res}}$, and the rms values of $u'_z$ and $ \theta'_\mathrm{res}$ scale with $\mathrm{Ra}$ in such a way that $\mathrm{Nu} \sim \mathrm{Ra}^{0.32}$;  without these corrections, $\mathrm{Nu} \sim \mathrm{Ra}^{1/2}$ in the turbulent regime.  For details refer to Pandey {\em et al.}~\cite{Pandey:2016a} and Pandey and Verma~\cite{Pandey:PF2016}.

\begin{figure}
\begin{center}
\includegraphics[scale=0.5]{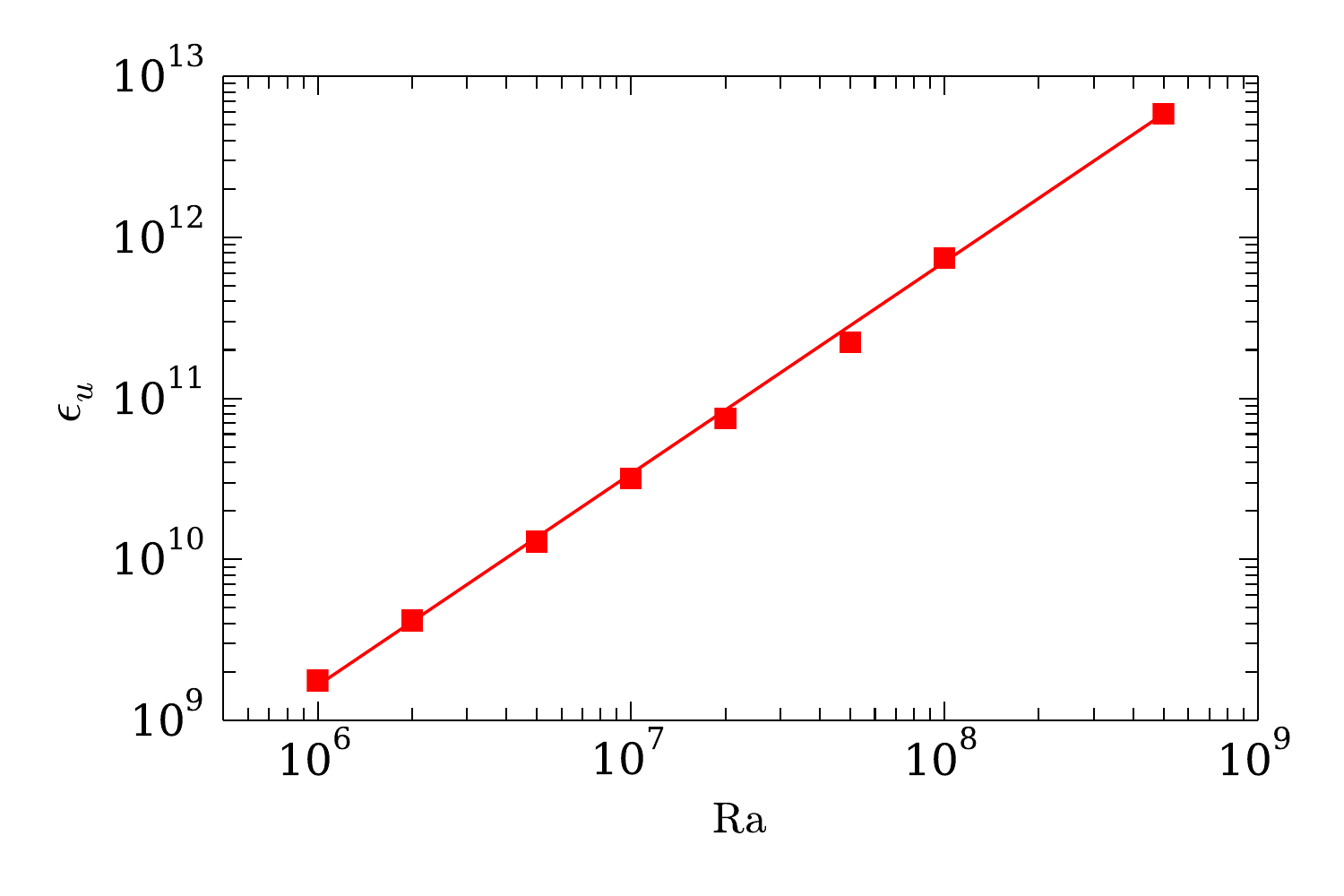}
\caption{A plot of the viscous dissipation rate $\epsilon_u$ vs.~$\mathrm{Ra}$.  The best fit curve is $\epsilon_u \sim \mathrm{Ra}^{1.32}$, indicating that $\epsilon_u \sim (U^3/d)\mathrm{Ra}^{-0.21}$ since $U \sim \mathrm{Ra}^{0.51}$.}
\label{fig:eps_u}
\end{center}
\end{figure}

In hydrodynamic turbulence, the viscous dissipation rate $\epsilon_u  \approx U^3/d$.  However this is not the case in RBC, primarily due to walls or boundary layers.  Using numerical data, Verma {\em et al.}~\cite{Verma:PRE2012} and Pandey {\em et al.}~\cite{Pandey:2016a}  have shown that $\epsilon_u \sim \mathrm{Ra}^{1.32}$ or
\be
\epsilon_u \sim (U^3/d)  \mathrm{Ra}^{-0.21}.
\ee 
See Fig.~\ref{fig:eps_u} for illustration for $\mathrm{Pr}=1$.  One of the exact relations of Shraiman and Siggia~\cite{Shraiman:PRA1990} yields
\begin{equation}
\epsilon_u  = \frac{U^3}{d}  \frac{\mathrm{(Nu-1)RaPr}}{\mathrm{Pe}^3}. \label{eq:eps_u1}
\end{equation}
Substitution of $ \mathrm{Pe} \sim \mathrm{Ra}^{0.51}$ and $\epsilon_u  \sim (U^3/d)\mathrm{Ra}^{-0.21}$ yields $\mathrm{Nu} \sim \mathrm{Ra}^{0.32}$.  These arguments show that the reduction of the viscous dissipation rate could be a reason for the deviation of the observed scaling $\mathrm{Nu} \sim \mathrm{Ra}^{0.32}$ from $\mathrm{Nu} \sim \mathrm{Ra}^{1/2}$ corresponding to the ultimate regime.

We conclude this section with a remark that the walls or the boundary layers affect the scaling of P\'{e}clet and Nusselt numbers significantly.

\section{Large-scale flow structures and flow reversals}
\label{sec:LSC}
The flow properties in the last two sections are related to the random nature of the flow.  It has been observed that coherent structures too play important role in the convective flow, and they have certain universal properties.  An interesting phenomena of RBC related to large-scale structures is  {\em flow reversals}.  Sreenivasan {\em et al.}~\cite{Sreenivasan:PRE2002}, Brown and Ahlers~\cite{Brown:JFM2006}, Xi and Xia~\cite{Xi:PRE2007}, and Sugiyama {\em et al.}~\cite{Sugiyama:PRL2010} observed  that the vertical velocity near the lateral wall switches sign randomly.   Deciphering how the reversals take place is an interesting puzzle, and it is not yet fully solved. In this section we  briefly describe the present status of the field.

It is believed that the flow reversals are caused by the nonlinear interaction among the large-scale structures of the flow.  For a closed cartesian box, these structures can be conveniently described by the small-wavenumber Fourier modes~\cite{Chandra:PRE2011,Chandra:PRL2013}.  This description is useful even for no-slip boundary conditions since the flow structures inside the boundary layers contribute to the large wavenumber modes.   For a cylindrical geometry, partial information about the flow structures can be obtained by computing the azimuthal Fourier modes corresponding to  the velocity field measured at various angles near the later walls~\cite{Brown:JFM2006,Xi:PRE2007,Mishra:JFM2011}.  Here we summarise the main results on the properties of flow reversals.
\begin{enumerate}
\item  During  a reversal, the amplitude of the most dominant large-scale mode vanishes, while that of the secondary mode rises sharply.   Chandra and Verma~\cite{Chandra:PRE2011,Chandra:PRL2013} reported that during a reversal in a unit cartesian box, the Fourier mode $(1,1)$ vanishes, while the mode $(2,2)$, corresponding to the corner rolls, become the most dominant mode~\cite{Chandra:PRE2011,Chandra:PRL2013}.  See Fig. 1 of Chandra and Verma~\cite{Chandra:PRE2011}.  This numerical result is consistent with the experimental results of Sugiyama {\em et al.}~\cite{Sugiyama:PRL2010}.  

 \item  The nature of dominant structures depends on the box geometry and boundary conditions. For example, for a box of size $2 \times 1$, under the no-slip boundary condition, $(2,1)$ and $(2,2)$ are the primary and  secondary modes respectively.  However, under the free-slip boundary condition, the corresponding modes are $(2,1)$ and $(1,1)$ respectively~\cite{Breuer:EPL2009,Verma:PF2015a}.
 
 \item Verma {\em et al.}~\cite{Verma:PF2015a} have constructed a group-theoretic argument to decipher  the reversing and non-reversing modes during a reversals.  The structure of the groups is related to the Klein group. 
 
\item Convection in a cylinder exhibit reversals that have similar behaviour as above.  Brown and Ahlers~\cite{Brown:JFM2006} termed such reversals as {\em cessation-led reversals}.  Note however that during a  cessation-led reversal, the secondary modes become significant, hence the kinetic energy does not vanish.  

\item Cylindrical convection exhibits another kind of flow reversals, called {\em rotation-led reversals}, in which the large-scale structure rotates azimuthally~\cite{Brown:JFM2006,Xi:PRE2007,Mishra:JFM2011}.  This rotation is due to the azimuthal rotation symmetry of the system. Such phenomena is also observed in a cylindrical annulus~\cite{Nath:ANE2013}. 
\end{enumerate}

The magnetic field reversals in dynamo, and the velocity field reversals in  Kolmogorov-flow also involve nonlinear interactions among the large-scale structures of the flow. Thus, these reversals share certain  similarities with the flow reversals of RBC.

\section{Summary}
\label{sec:summary}

In this short review we describe the spectral and large-scale properties of buoyancy-driven turbulence---stably-stratified flows and Rayleigh-B\'{e}nard convection.  A summary of the results covered in this review is as follows:

\begin{enumerate}
\item The stably-stratified turbulence (SST) is nearly isotropic for Froude number $\mathrm{Fr} \gtrapprox 1$.  Bolgiano~\cite{Bolgiano:JGR1959} and Obukhov~\cite{Obukhov:DANS1959} showed that for gravity-dominated flows ($\mathrm{Fr} \approx 1$), the kinetic-energy spectrum $E_u(k) \sim k^{-11/5}$.   Kumar {\em et al.}~\cite{Kumar:PRE2014} demonstrated this scaling using numerical simulations.

\item For $\mathrm{Fr} \gg 1$, SST exhibits Kolmogorov scaling, i.e. $E_u(k) \sim k^{-5/3}$, due to the dominance of the nonlinear term over the buoyancy.

\item For $\mathrm{Fr} \ll 1$, SST is quasi two-dimensional, and the kinetic-energy spectrum exhibits a combination of $k^{-5/3}$ and $k^{-3}$.

\item Turbulence in Rayleigh-B\'{e}nard convection (RBC) has strong similarities with the hydrodynamic turbulence, e.g, it exhibits constant energy flux and $k^{-5/3}$ energy spectrum in the inertial range. 

\item In RBC  turbulence, the pressure gradient accelerates the flow, while the buoyancy is balanced by the viscous dissipation.  This observation is consistent with the Kolmogorov-like phenomenology observed for RBC.

\item The aforementioned phenomenology of RBC turbulence is expected to work for other buoyancy-driven flows in which buoyancy feeds the kinetic energy.  Some of the examples of such flows are bubbly turbulence, non-Boussinesq thermally-driven flows in stars, turbulent buoyancy-driven exchange flows in a vertical pipe~\cite{Arakeri:CurrSci2000}, etc.

\item The scaling of the Reynolds and Nusselt numbers of RBC are well described by the models of Grossmann and Lohse~\cite{Grossmann:JFM2000, Grossmann:PRL2001, Grossmann:PRE2002, Grossmann:PF2004, Grossmann:PF2011}, and by the recent formula of Pandey {\em et al.}~\cite{Pandey:2016a}.
\end{enumerate}

We conclude this paper with a remark that several issues remain unresolved in buoyancy-driven turbulence, for example, the scaling of Nusselt number, the role of the boundary layer and bulk to  turbulence, existence of the ultimate regime.  High-resolution simulation, advanced experiments, and carefully modelling may resolve these outstanding questions in future.

%

\addcontentsline{toc}{section}{Acknowledgments}
\ack
We thank Stephan Fauve, J\"{o}rg Schumacher, Pankaj Mishra, Ravi Samtaney, Mani Chandra, Supriyo Paul,  Anando Chatterjee, and Jayant Bhattacharjee   for valuable discussions. Our numerical simulations were performed on  Cray XC40 ``Shaheen II" at KAUST supercomputing laboratory, Saudi Arabia. This work was supported by the research grants SERB/F/3279 from Science and Engineering Research Board, India, and PLANEX/PHY/2015239 from Indian Space Research Organisation, India.

\addcontentsline{toc}{section}{References}
{\bf References} \\


\end{document}